\documentclass[conference]{IEEEtran}
\usepackage[nocompress]{cite}

\usepackage{filecontents}
\usepackage{amsmath}
\usepackage{bbding}
\usepackage[nointegrals]{wasysym}
\usepackage{makecell}
\usepackage{xcolor}
\usepackage{multirow}
\usepackage{balance}
\usepackage{algorithm}
\usepackage{algorithmic}
\usepackage[algo2e]{algorithm2e}
\usepackage{comment}
\usepackage{url}
\usepackage{tabularx}
\usepackage{listings}
\usepackage[normalem]{ulem}
\usepackage{qtree}
\usepackage{booktabs}
\usepackage{color,colortbl,array}
\usepackage{textcomp}
\usepackage{graphicx}

\usepackage{subcaption}
\usepackage{longtable}

\usepackage{soul}
\usepackage{threeparttable}
\usepackage{arydshln} 
\usepackage{threeparttable} 
\usepackage{pifont}
\usepackage{hyperref}
\usepackage[numbers]{natbib}

\usepackage[utf8]{inputenc}

\usepackage{supertabular}
\usepackage {wasysym}
\usepackage{ltablex}
\usepackage{xurl}

\def\BibTeX{{\rm B\kern-.05em{\sc i\kern-.025em b}\kern-.08em
    T\kern-.1667em\lower.7ex\hbox{E}\kern-.125emX}}

\newcommand\geng[1]{{\color{purple}{\textbf{\{geng: {\em#1}\}}}}}

\newcommand\wmy[1]{{\color{blue}{\textbf{\{wmy: {\em#1}\}}}}}
\newcommand{\ww}[1]{{\color{orange}#1}}

\newcounter{finding}
\newcommand{\finding}[1]{
\vspace{3pt}
\noindent
\framebox{
\begin{minipage}[b]{0.96\linewidth}
\noindent \textbf{Finding \Roman{finding}}: \textit{#1}
\stepcounter{finding}
\end{minipage}}
\vspace{3pt}}

\newcommand{\eg}{{\it e.g.,}\xspace}

\newcommand{\etc}{{\it etc.}\xspace}
\newcommand{\ie}{{\it i.e.,}\xspace}

\newcommand{\ignore}[1]{}

\newcommand{\engine}{device search engine\xspace}
\newcommand{\engines}{device search engines\xspace}
\newcommand{\Engines}{Device search engines\xspace}
\newcommand{\scanip}{{\it ScanIP}\xspace}
\newcommand{\scanips}{{\it ScanIPs}\xspace}
\newcommand{\ipmirror}{{Mirror Service}\xspace}
\newcommand{\ipmirrors}{{Mirror Services}\xspace}

\newcommand{\evil}{\multirow{1.3}{*}{\includegraphics[height=1em]{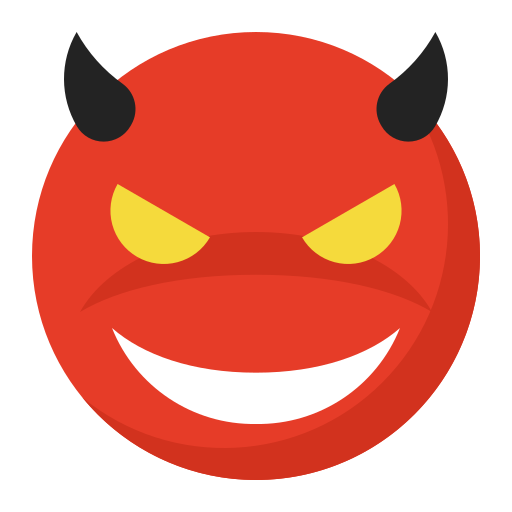}}}
\newcommand{\neutral}{\multirow{1.3}{*}{\includegraphics[height=1em]{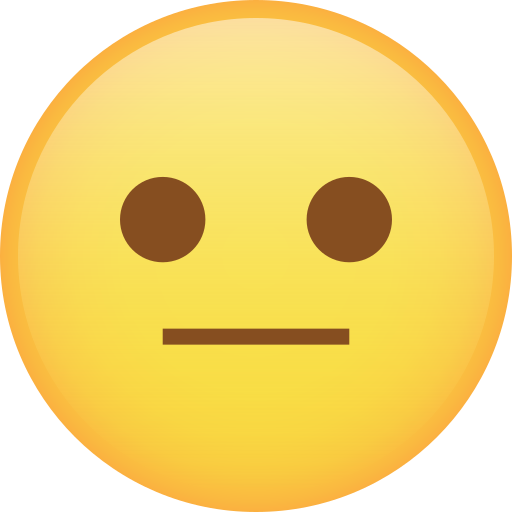}}}
\newcommand{\benign}{\multirow{1.3}{*}{\includegraphics[height=1em]{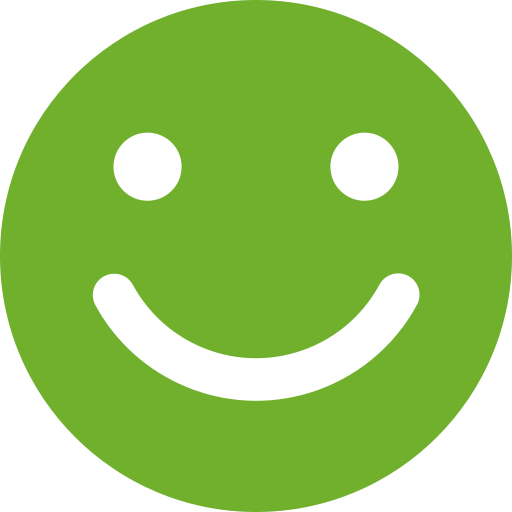}}}

\newcommand{\noteEvil}{\raisebox{-0.5ex}{{\includegraphics[height=1em]{fig/devil.png}}}}
\newcommand{\noteNeutral}{\raisebox{-0.5ex}{{\includegraphics[height=1em]{fig/flat.png}}}}
\newcommand{\noteBenign}{\raisebox{-0.5ex}{{\includegraphics[height=1em]{fig/smile.png}}}}

\newcommand{\del}[1]{\textcolor{red}{\sout{#1}}}

\begin{document}

%\newpage

\title{
% Device Search Engines: A Double-Edged Sword in Network Security and Privacy
Revealing the Black Box of Device Search Engine: Scanning Assets, Strategies, and Ethical Consideration \vspace{-0.15in}
}
% 
% 逆向设备搜索引擎的黑盒
% understanding 测绘引擎的策略

\author{
    \IEEEauthorblockN{
        Mengying Wu\IEEEauthorrefmark{2}\IEEEauthorrefmark{1},
        Geng Hong\IEEEauthorrefmark{2}\IEEEauthorrefmark{1},
        Jinsong Chen\IEEEauthorrefmark{2}, 
        Qi Liu\IEEEauthorrefmark{2},
        Shujun Tang\IEEEauthorrefmark{3}\IEEEauthorrefmark{4},
        Youhao Li\IEEEauthorrefmark{3}, 
        Baojun Liu\IEEEauthorrefmark{4}, \\
        Haixin Duan\IEEEauthorrefmark{4}\IEEEauthorrefmark{5} and
        Min Yang\IEEEauthorrefmark{2}
    }
    \IEEEauthorblockA{
        \IEEEauthorrefmark{2}Fudan University, China,
        \{wumy21,jschen23,qiliu21\}@m.fudan.edu.cn,
        \{ghong,m\_yang\}@fudan.edu.cn
    }
    \IEEEauthorblockA{
        \IEEEauthorrefmark{3}QI-ANXIN Technology Research Institute, China,
        liyouhao@qianxin.com
    }
    \IEEEauthorblockA{
        \IEEEauthorrefmark{4}Tsinghua University, China,
        tsj23@mails.tsinghua.edu.cn,
        \{lbj, duanhx\}@tsinghua.edu.cn
    }
    \IEEEauthorblockA{
        \IEEEauthorrefmark{5}Quancheng Laboratory, China
    }
}

\IEEEoverridecommandlockouts
\makeatletter\def\@IEEEpubidpullup{6.5\baselineskip}\makeatother
\IEEEpubid{\parbox{\columnwidth}{
		Network and Distributed System Security (NDSS) Symposium 2025\\
		24-28 February 2025, San Diego, CA, USA\\
		ISBN 979-8-9894372-8-3\\
		https://dx.doi.org/10.14722/ndss.2025.241924\\
		www.ndss-symposium.org
}
\hspace{\columnsep}\makebox[\columnwidth]{}}

\maketitle

\renewcommand{\thefootnote}{\IEEEauthorrefmark{1}}
\footnotetext{These authors contributed equally to this work.}
\renewcommand{\thefootnote}{\arabic{footnote}}

\begin{abstract}

% 设计了framework 收集了ip 理解策略
% 认识了协议识别
% ethical
In the digital age, device search engines such as Censys and Shodan play crucial roles by scanning the internet to catalog online devices, aiding in the understanding and mitigation of network security risks. While previous research has used these tools to detect devices and assess vulnerabilities, there remains uncertainty regarding the assets they scan, the strategies they employ, and whether they adhere to ethical guidelines.
% 理解策略对哪些人有帮助

This study presents the first comprehensive examination of these engines' operational and ethical dimensions. We developed a novel framework to trace the IP addresses utilized by these engines and collected 1,407 scanner IPs. 
By uncovering their IPs, we gain deep insights into the actions of \engines for the first time and gain original findings.
By employing 28 honeypots to monitor their scanning activities extensively in one year, we demonstrate that users can hardly evade scans by blocklisting scanner IPs or migrating service ports. Our findings reveal significant ethical concerns, including a lack of transparency, harmlessness, and anonymity. Notably, these engines often fail to provide transparency and do not allow users to opt out of scans. Further, the engines send malformed requests, attempt to access excessive details without authorization, and even publish personally identifiable information(PII) and screenshots on search results.
% thereby breaching principles of ethical conduct such as transparency, respect, and harmlessness.\wmy{?} 
These practices compromise user privacy and expose devices to further risks by potentially aiding malicious entities. This paper emphasizes the urgent need for stricter ethical standards and enhanced transparency in the operations of device search engines, offering crucial insights into safeguarding against invasive scanning practices and protecting digital infrastructures.

\end{abstract}

\IEEEpeerreviewmaketitle

\section{Introduction}
% 是什么，正常人拿它干什么
Device search engines like Censys\cite{durumeric2015search} and Shodan\cite{shodan} scan the entire Internet to catalog online devices, maintaining up-to-date records of hosts and services within the public IPv4 address space.
These engines are crucial for helping engineers understand network security risks by offering comprehensive and robust data support. Researchers frequently utilize device search engines to build a data-driven view of device landscape and surging vulnerabilities impact. For instance, prior studies employed these engines to collect data on resident IP addresses~\cite{mi_resident_2019}, electric vehicle charging management systems~\cite{nasr_chargeprint_2023}, and insecure industrial control systems (ICS)~\cite{sasaki_exposed_2022}.
 % for remote management

% 引出问题 工作机制不清晰 也不知道隐私和安全问题

% 强调其能力会被恶意攻击者滥用
% On the one hand, 
Attackers can abuse the powerful scanning capabilities of such engines to identify vulnerable devices and establish zombie networks for malicious activities like cryptocurrency mining~\cite{xulu}. It is estimated that the over-collection of data by Shodan-like services led to a loss of approximately \$3.86 million in 2020 alone~\cite{dangerous-shodan}.
Moreover, it remains uncertain whether these engines consider ethical implications while striving to provide competitive network assessment reports.
% On the other hand, despite their advanced scanning techniques, it remains uncertain whether these engines adhere to their ethical commitments.
%
Users who care about security and privacy have started to take action, including reporting abusive scanning IPs to AbuseIPDB~\cite{abuseipdb}, a public IP blocklist, and moving services from default ports to other ports.
To the best of our knowledge, there has been limited effort to thoroughly examine the operational strategies, and potential ethical violations associated with these engines.

% Despite advances in utilizing device search engines, their operational strategies and ethical concerns remain largely unexplored,

% this paper干啥了
To fill this gap, this paper presents the first measurement study on the working strategies of \engines and reveals their potential aggressive behavior and privacy issues. 
Our study is driven by the following research questions (RQs):

\begin{itemize}
    % \item[RQ1] What is the scanning strategy of \engines? \wmy{是否有可能定个blacklist，不可能}
    \item[RQ1] Can users block the IPs of \engines to avoid being scanned?
    % \item[RQ2] How does a \engine identify protocols on a host?
    \item[RQ2] Can users migrate service ports to avoid being scanned?
    \item[RQ3] Will the scanning of the \engine introduce any security or privacy concerns to the services being scanned? 
\end{itemize}

% 用terminal做一个ipmirror的图，然后search engine的效果

\begin{figure}[t]
    \centering
    \includegraphics[width=\linewidth]{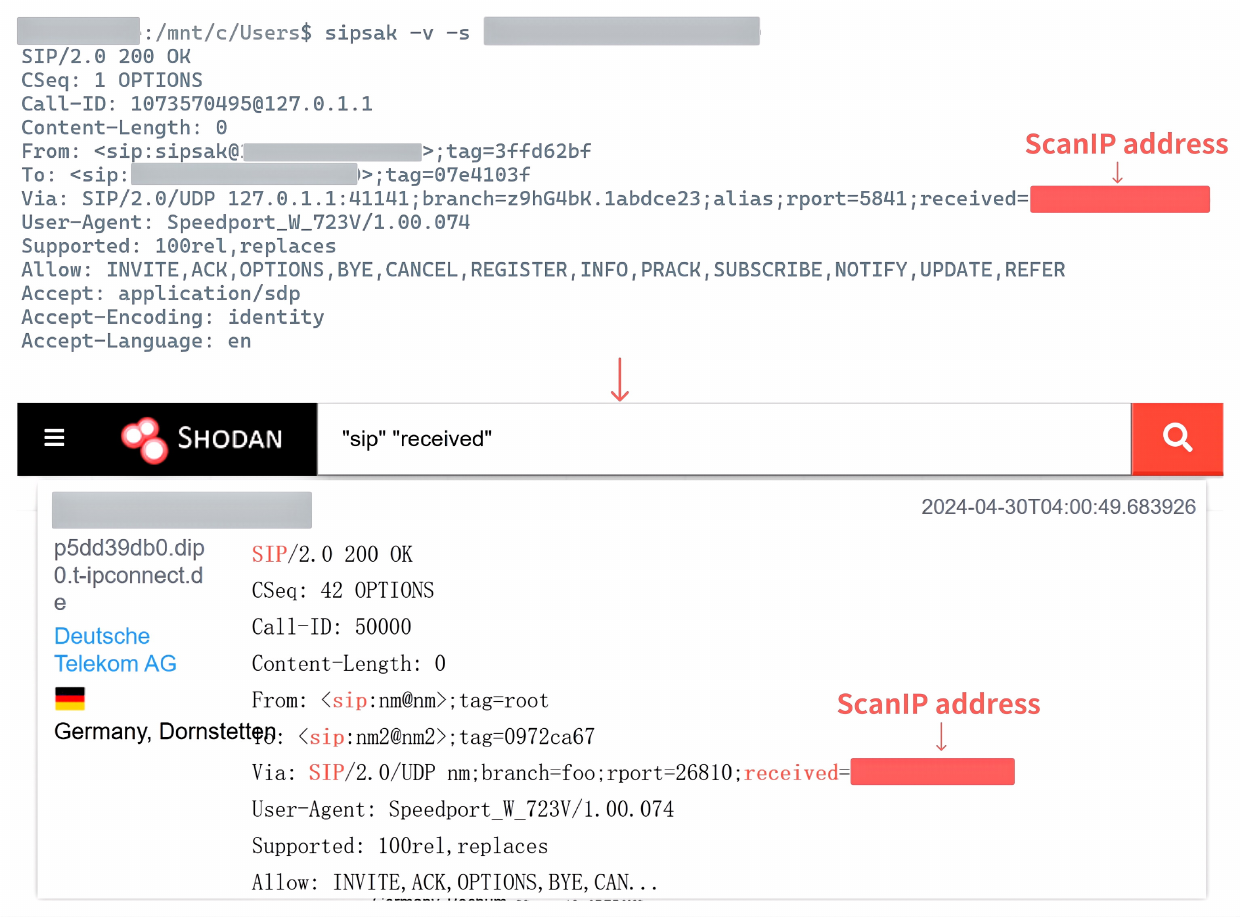}
    \caption{IP Mirror Service. Interacting with a SIP server by \texttt{sipsak -v -s \{IP\}} will reply with the sender's IP. When querying \texttt{\{"SIP" "received"\}} in Shodan, its \scanip is shown under the red mark.}
    \label{fig:ipmirror}
\end{figure}

% challenge 1
\noindent\textbf{Challenges.}  
% 前人针对web bot是怎么发现的，但是测绘引擎xxx
Prior works~\cite{sun2010ethicality, li2021good} used User-Agent (UA) headers to identify traffic from search engines, as the Robots Exclusion Protocol~\cite{robot_exclusion} is commonly adopted by these engines. However, to profile the behavior of \engines, \textit{the main difficulty lies in differentiating \engine scanning activities from others}, largely due to most \engines' hesitance to disclose their IP lists and only web requests containing UA as labels.
%, hindering our tracking efforts.
% 
Even getting their IPs, no ready-made system is available to comprehensively understand scanning strategies and detect potential ethical violations.

\noindent\textbf{Insights.} 
% \geng{there is a gap between profile behavior and find IP then set up a honeypot.}
% To profile the behavior of \engines, we first  identified a unique service, termed \ipmirror, where responses contain the IP of the request sender
% insight 1
To find the IPs of \engines, we identified a unique service, termed \textit{\ipmirror}, where responses contain the IP of the request sender, as shown in Figure~\ref{fig:ipmirror}. Network services may include the visitor's IP for debugging, error prompting, or log metadata purposes. For example, when communicating with SIP (Session Initiation Protocol)~\cite{rfc3261},
% is a communication protocol utilized for establishing, modifying, and terminating multimedia sessions over a network. 
the visitor’s IP address is shown when a proxy receives the request from a different address than the one specified in the header. 
% Within SIP responses, the originating IP address of the communication is recorded, facilitating the establishment of a communication link with that specific IP.\wmy{different with section III}
\textit{When \engines scan those services,
% and show the response in the engine, 
their IP addresses (\scanip) are inevitably logged.}
% get reflected.
%

Getting the \scanips helps to distill engines' action. Specifically,  we use honeypots to capture in-depth behavior effectively.
%, as they are controlled environments that allow for observation and analysis.
% \del{ like previous works do~\cite{li2021good}.}
% \wmy{little abstract, did not say what previous work do} 
While simulating every device type for monitoring scans is impractical, we focused on IoT devices due to their ubiquity. Device search engines are widely used for discovering IoT devices, each offering complex services for identification. This targeted approach allowed us to capture in-depth behaviors of \engines effectively.

\noindent\textbf{Our work.}
We develop a systematic framework to retrieve \ipmirrors from engines' search results and collect \scanips based on the \ipmirror banners. We applied our framework to four engines 
% with batch-automatable search API capabilities
: Censys~\cite{censys_search}, Shodan~\cite{shodan}, FOFA~\cite{fofa}, and ZoomEye~\cite{zoomeye}. 
% Furthermore, to address our RQs, we enhanced our honeypots with three specific designs featuring dynamic responses across multiple paths.
% \geng{please give a more intuitive description of our honeypot, not the DETAIL of our honeypot} to address our RQs,
We also deployed honeypots to learn scan strategies and evaluate their ethical consideration of scanning.
Through our innovative methodology, we gained deep insights into the actions of \engines for the first time, ensuring that our findings are original contributions to the field, not mere reiterations of publicly available information from the \engines themselves.
% design functional/multi-faceted/multi-dimensional
% \ww{Furthermore, we design sniffer honeypots to learn engines' scan strategy, and web honeypots that emulate IoT devices and trackable links to attract deeper scanning sessions that can evaluate ethical behaviors.}
% Furthermore, to address our RQs, we design a full-port closed honeypot, a popular-port open honeypot, and a web honeypot. 
% We enhanced the web honeypot with IoT device emulation, privacy paths, and dynamic trackable links, 
% To mitigate the impact of \scanip ownership changes and potential \ipmirror service downtime, we limited our analysis to data records in \engines collected between March 2023 and March 2024.

\noindent\textbf{Results.} Using data records in \engines collected between March 2023 and March 2024, we collected 106,132 \ipmirrors and 1,407 \scanips. FOFA has the most \scanips (665), followed by ZoomEye (166), Censys (140), and Shodan (91). 
% \footnote{All \scanips we collected are IPv4 addresses.}
% \geng{why highlight ipv4 in introduction?}\wmy{if reviewer ask ipv6?}

% 
\ignore{
During this process, we found that except for Censys, which discloses the \scanip in each record\geng{scan banner?}, the other \engines are deliberate about concealing the \scanip in their records. Particularly, Shodan replaces the \scanips with multicast addresses (224 IP segment)\cite{rfc5771}, leaving no trace, and assigns unique multicast addresses for each \scanip.
}

% 我们看到都在换ip，是否在blacklist里，所以我推断xxx
% 我们发现不同测绘引擎在端口上有不同的倾向，比如xxx，可能跟产品特点各有特征
\noindent\textbf{Scan Strategy.} We deployed 28 honeypots across the different countries and captured 7.4 million requests from 839 \scanips from March 2023 to March 2024, totaling 4.6GB of raw logs. 
% Through traffic analysis, we discovered that engines often prefer using IPs from their own countries. 
We found that FOFA and ZoomEye did not use fixed scanning IPs, with \textit{FOFA typically rotating its \scanips every three months}. As 665 IPs we found are reported abusive in AbuseIPDB by users, the rotation may aim to avoid being blocklisted by users (see Section~\ref{sec:landscape}).
% however, these trashed IPs have been exploited by attackers for malicious scanning activities, confirmed by FOFA.
% 
The port preference among the engines differs. 
% ZoomEye preferred scanning ports that have a high risk for potential DDoS attacks such as game server(27015) and BitTorrent(6881), while others preferred scanning commonly used service ports like HTTPS(443), SSH(22), and Telnet(23)(see Section~\ref{sec:landscape}).
ZoomEye primarily scanned high-risk DDoS ports, while other engines focused on common service ports like HTTPS, SSH, and Telnet (see Section~\ref{sec:port_scan}).

% Despite having hundreds of \scanips, no engine manages to scan all ports within a year, and even port 443 data cannot be updated daily in FOFA and ZoomEye (see Section~\ref{sec:landscape})\wmy{need confirm}\geng{revise according to our discussion please.}.

\noindent\textbf{Protocol Identification.} For identifying services on open ports, we found that \textit{\engines probe services not only on default ports but also on neighbor ports} (see Section~\ref{sec:service}), this indicates that users who migrate the ports of services cannot conceal the service being indexed by device search engine effectively. For example, RDP is probed on ports 3388 to 3390.
% This practice indicates that changing to neighboring ports does not effectively conceal the service. 
When engines fail to identify the default protocol, they adopt fallback strategies: most prefer HTTP and HTTPS, while FOFA switches to FTP and ZoomEye to RDP.

\noindent\textbf{Ethical Scanning.} 
Various countries have enacted cybersecurity laws~\cite{NIS2,CCPA,ChinaCyberSecurityLaw} and personal information privacy laws~\cite{gdpr,CFAA,ChinaDataSecurityLaw} to safeguard people's rights. Guidelines also exist to regulate scanning and crawling behaviors. However, scanning and indexing device and service information may violate the principles of transparency, harmlessness, and anonymity.
% 
% However, the implementation of these laws in \engines lacks clear standards. 
To assess the potential violation (see Section~\ref{sec:ethics}), we summarize the guidelines based on best practices from popular scanning tools~\cite{durumeric_zmap_2013,durumeric2015search,onyphe-standard}, crawler standards~\cite{rfc9309,guideline_robot}, and ethical principles~\cite{dittrich2012menlo}. 
% we discovered multiple ethical problems in the \engines. 
% While ethical standards for device search engines lack legal benchmarks, there are best practices. 

As for transparent scanning, \engines should inform individuals about who is collecting their data, why it is being collected, and how to opt out. Notably, \textit{users cannot discern whether scans originate from FOFA or ZoomEye} through IP homepages, WHOIS, Reverse DNS, or public listings. Apart from Censys, none provide opt-out options, and most conceal their identity in the User-Agent. 
Additionally, we observed that Censys does not adhere to its recommended practice~\cite{durumeric2015search} of explaining the scanning purpose on every probe.

For harmless scanning, \engines should only send standard requests and access public resources. However, we observed that they send \textit{malformed requests, attempt unauthorized data collection}, and exploit vulnerabilities, risking user privacy and security.
In our investigation of 12 popular services, all four engines excessively attempted anonymous logins, retrieved system details, and enumerated database contents, exposing 214,862 Redis hosts and 135,599 FTP services that lack authentication and are vulnerable to arbitrary access.

% Publishing unanonymized sensitive data collected from Internet assets also violates privacy guidelines.
%
% \ww{Improper display of personal information can result in privacy leakage.}
% and re-identification risk, which help the privacy trafficking industry.}\geng{why mention re-identification and trafficking}\wmy{emphasize the consequence}
For anonymity, we witness \engines publishing unanonymized sensitive data in search results, including PII~(name, email, avatar, screenshots, \etc) and database entries.
Specifically, Specifically, Shodan lists data entries for 68,543 Redis hosts, while FOFA and ZoomEye publish 145,310 database indices of Elasticsearch.
Notably, 904,303 snapshots of IP cameras and screenshots of remote desktops are displayed in Shodan, \textit{as a paid service}.

% Additionally, these engines do not handle the presence of sensitive devices with privacy measures, which can aid attackers in exploring user privacy.

\noindent\textbf{Contributions.} This paper makes the following contributions:

\noindent$\bullet$ We proposed a semi-automated framework for discovering services that can reflect \scanips of \engines and uncover 1,407 \scanips.

% \noindent$\bullet$ We conduct the first comprehensive analysis of the scan strategy of \engines, one of the most critical infrastructures in the security field.
\noindent$\bullet$ We conduct the first comprehensive analysis of the scan strategy of \engines, demonstrating that users cannot evade scans by blocklisting scanner IPs.

\noindent$\bullet$ We unveil how \engines identify protocol on ports, offering insights into how users can hide their services.

\noindent$\bullet$ We conducted an ethical analysis of \engine scanning behaviors, uncovering instances where engines conceal their identities, engage in unauthorized access, and expose user camera interfaces.

\ignore{
\noindent$\bullet$ We release our \ipmirror list and code~\cite{DSEdataset} to \geng{should we public such data? the propose ``help identifying" not convincing me.} help user identify \engines.
}

\section{Background and Related Work}
% \subsection{Device Search Engine Technique}
% 以事件形式串，不要以作者形式串
\begin{table*}[t]
    \centering
    \caption{ \scanips across different services of \engines in the preliminary study. $\Circle$ represents the \scanips are hid, $\CIRCLE$ represents the \scanips are shown in a standard form, $\RIGHTcircle$ represents the \scanips are shown in a reverse form, $\LEFTcircle$ represents the \scanips are encoded in URL, and - represents we did not find records containing that attribute for the specified service.}
    \label{tab:leak_protocol}
    % \resizebox{\linewidth}{!}{
    \begin{tabular}{c c c c c c c c}
        \toprule
        \multirow{2}{*}{\textbf{Engine}} & \multirow{2}{*}{\textbf{Country}} & \multirow{2}{*}{\textbf{Year}} & \textbf{HTTP}   & \textbf{MySql}   & \textbf{SIP} & \textbf{SMTP} & \textbf{HTTP} \\
        \cline{4-8} 
        & & & \multirow{1.3}{*}{\textbf{X-Forward-For}} & \multirow{1.3}{*}{\textbf{ERR\_HOST}} & \multirow{1.3}{*}{\textbf{Received}} & \multirow{1.3}{*}{\textbf{No Valid PTR}} & \multirow{1.3}{*}{\textbf{Location}} \\
        \midrule
        Shodan\cite{shodan}  & USA & 2009 & $\CIRCLE$ & $\Circle$ & $\Circle$ & $\RIGHTcircle$ & $\LEFTcircle$\\
        ZoomEye\cite{zoomeye} & China &2013 & $\CIRCLE$ & $\CIRCLE$ & $\CIRCLE$ & $\RIGHTcircle$  & $\LEFTcircle$\\
        Censys\cite{censys_search} & USA & 2015 & $\CIRCLE$ & $\CIRCLE$ & - &$\RIGHTcircle$ & $\LEFTcircle$ \\
        FOFA\cite{fofa}  & China & 2015 & $\CIRCLE$ & $\Circle$ & $\Circle$ & $\RIGHTcircle$  & $\LEFTcircle$\\
        BinaryEdge\cite{binaryedge} &Switzerland & 2015 & $\CIRCLE$ &$\CIRCLE$ & $\CIRCLE$ &$\RIGHTcircle$   & $\LEFTcircle$ \\
        Netlas\cite{netlas} & Armenia &2021 & $\CIRCLE$ & $\CIRCLE$& - & -  & - \\
        Hunter \cite{hunter} & China & 2021 & $\CIRCLE$ &$\CIRCLE$ & $\CIRCLE$ &$\RIGHTcircle$   &- \\
        \bottomrule
    \end{tabular}
    % }
    % attributes or error codes (second row) of  (first row) 
\end{table*}

% https://cloud.tencent.com/developer/article/1078379 -> ZoomEye
% https://en.wikipedia.org/wiki/Shodan_(website) -> Shodan
% https://docs.netlas.io/changelog/ -> Natlas
% https://hunter.qianxin.com/home/changelog -> Hunter

Facing the rising requirement of internet analysis, there has been an increasing number of device search engines in recent years, as listed in Table~\ref{tab:leak_protocol}.
These engines are specialized scanning tools that index information about internet-connected devices. They provide Internet threat intelligence, consisting of device types, running services, and potential vulnerabilities. This data is utilized by security researchers, network administrators, and even cyber attackers to locate weaknesses.
% or understand network setups.

These engines collect detailed asset records, including IP addresses, ports, timestamps, geographical locations, and banner content. They also offer advanced features such as service version labeling, protocol identification, honeypot detection, certificate analysis, and vulnerability detection.

To facilitate result querying, \engines maintain up-to-date snapshots of hosts and offer a user interface (UI) and APIs. They typically index responses and develop engine-specific search syntax, allowing users to filter and access targeted assets effectively.
% To facilitate results querying, \engines generally index the responses banners and develop engine-specific search syntax.

% Device search engines are search engines for discovering online devices. Common \engines include Censys, Shodan, FOFA, and Zoomeye.
% They perform scans across the entire internet, offering a cloud-based service that not only maintains an up-to-date snapshot of hosts and services operating within the public IPv4 address space but also publishes these data through search engines and APIs.

Previous \engine research has primarily focused on developing scanning techniques and toolchains, as well as analyzing internet behavior facilitated by these tools.

\noindent \textbf{Scanning Tools.}
Internet-scale scanning tools like nmap~\cite{nmap} and ZMap~\cite{durumeric_zmap_2013} are fundamental components of \engines, used to initiate host discovery within the address space.
% 在多长时间内完成多快的扫描
Censys~\cite{durumeric2015search}, which employs ZMap to conduct single-packet host discovery scans across the IPv4 address space in 45 minutes, effectively mapping out reachable hosts. 
Other tools, such as IRLscanner~\citep{leonard2010demystifying} and MASSCAN~\cite{graham2014masscan}, can scan the entire Internet in under five minutes, while Zippier ZMap~\citep{adrian2014zippier} dramatically improves scanning speed to 4.5 minutes by parallelizing address generation and utilizing zero-copy NIC access.

% 用什么方法搞的更快
% Internet-scale scanning tools, such as nmap~\cite{nmap} and Zmap~\cite{durumeric_zmap_2013}, are the backbone of \engines. These tools are typically used to initiate the scanning process, enabling \engines to discover hosts within the address space.
% Taking Censys\cite{durumeric2015search} as an example, it initially utilizes Zmap for single-packet host discovery scans across the IPv4 address space. Subsequently, application-layer handshakes are conducted on identified open hosts, facilitating the measurement of various aspects of their service configurations.
% Other scanning tools including IRLscanner~\citep{leonard2010demystifying}, MASSCAN~\cite{graham2014masscan}, Zippier ZMap\citep{adrian2014zippier}, which dramatically enhanced scanning speed by a factor of ten. 
% \citet{izhikevich2022predicting} developed a system for predicting IPv4 services across all ports. 

% These contributions provide researchers with a diverse resource and ethical considerations for analyzing internet behavior. The ethical dimension highlighted in these studies has set the standards for \engine, a crucial aspect that our work further explores.

\noindent \textbf{Behavior Analysis with Device Search Engines.}
Researchers have used the indexed results from \engines to detect potential vulnerabilities and assess their severity, particularly in IoT devices.
% As Internet-scale network surveys collect data by probing large subsets of the public IP address space, researchers usually utilize \engines to detect devices and understand the component coverage and vulnerability impact range of IoT devices. 
For instance, prior works utilize \engines to collect resident proxy IPs~\cite{mi_resident_2019}, electric vehicle charging management systems~\cite{nasr_chargeprint_2023} and search for Mirai bots from HTTPS, FTP, SSH, Telnet, and CWMP~\cite{antonakakis_understanding_2017}, determining the types of infected devices~\cite{cetin_cleaning_2019}.
\citet{srinivasa2021open} unveiled 1.8 million misconfigured IoT devices without authority that may be exploited to perform large-scale attacks, \citet{sasaki_exposed_2022} detected 890 insecure ICS devices in Japan via their WebUI and discovered 13 0-day vulnerabilities. These works highlight the significant risk of over-sharing device information.

% 不关键 可以炸掉并起来
% 发现没办法跟前面的并起来 直接炸了

%%% 下面这一段话报错了 先注释掉了
% \del{
% \noindent\textbf{Internet Scanning Analysis.} \citet{durumeric2014internet} delved into port scanning activities across the internet based on network telescope, analyzing their scale, origins, and impact on network security. \citet{bano2018scanning} developed a comprehensive methodology for concurrent IPv4 scans using various protocols, analyzing nuanced interpretations of probe responses to improve active Internet measurement studies. \citet{quan2011detecting} introduced a method for detecting and analyzing internet outages by actively probing entire IP address blocks. \wmy{split and put somewhere}
% }

% Unlike these works, we conducted the first systematic study of the device search engines themselves, revealing their scanning strategies, service identification methods, and ethical violations.

\noindent\textbf{Bot Analysis.}
The most relevant works of ours are \cite{sun2010ethicality}, \cite{li2021good}, \cite{BODENHEIM2014114}, and \cite{zhao2020large}. 
\citet{sun2010ethicality} first measured web crawler ethicality and found most search engines respect robots.txt but misinterpret certain rules, while \citet{li2021good} 
% characterized automated browsing activity, 
uncovered the behavior and features of bots, particularly exposing the extensive activity of malicious bots. However, their reliance on user agents can not tell the behavior of \engines. 
\citet{BODENHEIM2014114} evaluated Shodan's indexing and querying capabilities on ICS, while \citet{zhao2020large} evaluated the vulnerability surface of IoT devices and utilized 60 days to learn the scanning period of engines. Both used records from a few servers (four and seven) to analyze engine scans on IP level.
In contrast, our paper introduces a method to discover mirror services reflecting \scanips, which hasn't been reported before. Armed with these unique viewpoints, we are able to analyze previously unknown device search engine assets for the first time, analyze scanning strategies using a one-year dataset, and conduct an ethical analysis, revealing unethical practices.
% \citet{BODENHEIM2014114} evaluated Shodan's ability to search PLC and proposed a potential method to defend against Shodan.
% \citet{zhao2020large} utilize \engines and show that N-days vulnerability is seriously endangering IoT devices. 
% % They also studied the responding time and IP-level scanning period of 5 \engines for a new host, however, they only used 7 servers for 60 days and only found Shodan is the only one that scans all the servers more than once, leading to a big bias on their result. 
% In contrast, our paper makes a systematic analysis of \engines, looking inside into the scanning asset,  scanning strategies, as well as their ethical concerns.

% 之前遗漏了什么
Even though the evolution of network-level scanning techniques has accelerated the ability of \engines to index Internet assets, questions remain regarding the ethics of their scanning practices. 
Can users blocklist their IPs or hide the services to avoid their scanning?
Do \engines conduct ethical scanning? 
Do they give users any ways to originate their scanning?
% Do they clearly explain the purpose of their massive scans? 
There is concern about whether their scanning would harm devices or expose hidden vulnerabilities and sensitive data, requiring further study.
\section{Preliminary Study}
To understand the behaviors of device search engines, the main challenge lies in differentiating the actions of these engines from other bots or scanners, since most device search engines did not publish their ScanIP lists nor announce in User-Agent when accessing web services.
% Therefore, it is crucial to identify 
This section introduces the \ipmirror, a service that can contain the requester’s IP address, and demonstrates how it provides us with an opportunity to analyze \engines.

% % \ww{We try to differentiate the actions by identifying the IP addresses (\scanip) of device search engines, extracting from their public records.}\geng{remove?}
% 我们对网站内容分析找到ip地址

\subsection{\ipmirror}
% \noindent \textbf{Response of services.} 
% The banner serves as a record of all plaintext responses from services, providing a window for users to understand the nature of a given service.
In network services, it's common for responses to include the IP information of the request sender, a phenomenon we refer to as ``\ipmirrors''.
These services may include the visitor's IP address in their responses for various reasons in design, such as debugging, error message, or log metadata.
% 有些协议在设计实现的时候为了报错、记录交互过程
For example, MySQL~\cite{mysql_err} responds to illegal connection attempts by notifying the attempting IP address that it does not have permission to connect to the server, showing \textit{``Host \{IP\} is not allowed to connect to this MySQL server''}.
Similarly, SIP (Session Initiation Protocol)~\cite{rfc3261}, a communication protocol used to establish, modify, and terminate multimedia sessions across networks, reveals the sender's IP when a User Agent (UA) or proxy receives a request from a different address than the one specified in the top \texttt{Via} header field.

% We named these services, which display the originating IP in their responses, ``\ipmirror''. 

\subsection{\ipmirror in Device Search Engine}\label{sec:preliminary}

Typically, \ipmirrors do not compromise security assumptions, as only the request receiver can log the sender's IP. However, the situation changes when \engines scan these \ipmirrors and display the services' responses. This inevitably exposes \scanip and also provides us with an opportunity to analyze the behavior of \engines.

% scanning results, supporting protocols, IP formatting, etc.; place a big/medium table here. methodology
To systematically survey the \ipmirrors in \engines, we identified 13 device search engines by using keywords such as ``cyber asset search engine'' and ``device search engine'' in search engines. We successfully registered accounts and accessed device data from seven of them. We manually inspected their search results concerning web services, MySQL~\cite{mysql}, SIP~\cite{rfc3261}, and SMTP~\cite{smtp_rfc}, checking whether and how their \scanip is presented.

% \begin{figure}[t]
%     \centering
%     \includegraphics[width=0.8\linewidth]{fig/snapshot_for_preliminiary.pdf}
%     \caption{The search result of querying \texttt{service:"SIP" +banner:"received"} in ZoomEye, which shows its \scanip under the red mark. \wmy{change to shodan}}
%     \label{fig:snapshot}
% \end{figure}

Table~\ref{tab:leak_protocol} shows that \ipmirrors are widely scanned and logged by \engines.
By searching engine records with specific \ipmirror traits (\eg \texttt{service:"SIP" + banner:"received"} for the SIP protocol in ZoomEye) 
and employing regular expressions to match IPv4 formats, IP addresses can be uncovered.
% within each engine's search results \geng{? why}\wmy{?}

% By combing through records of hosts exhibiting \ipmirror characteristics (such as using the SIP protocol and including the ``received='' keyword in responses) and employing regular expressions to match IPv4 address formats, IP addresses can be identified within each engine's search results, as shown in Figure~\ref{fig:snapshot}.
%
% Consequently, we can leverage these \ipmirrors to discover the \scanips of various \engines in a bulk manner. Given the critical role of banner information in revenue generation for engines, it is improbable that they would disable this service to evade detection of their \scanips.

\noindent \textbf{Formats of IPs.}
We found that IPs can be reflected in three formats—standard, reverse, and encoded—based on different service designs and requirements, as shown in Table~\ref{tab:leak_protocol}.

\begin{itemize}
    \item \textbf{Standard IP}, an IPv4 address represented using dotted-decimal notation, such as ``1.2.3.4''.
    \item \textbf{Reverse IP}, \ie \textit{``4.3.2.1.in-addr.arpa''}, is utilized in reverse DNS queries to find the domain name associated with an IP. Besides, SMTP servers may raise exceptions with the sender's IP in reverse form if no valid PTR (Pointer) record is found.
% in URL encoding, the dot (.) is converted to ``\%2E''. Thus,
    \item \textbf{URL encoding IP}, ensures that special characters are converted in URLs, preventing conflicts with URL structure and syntax. For instance,  the standard form IP \textit{``1.2.3.4''} becomes \textit{``1\%2E2\%2E3\%2E4''}. This encoding is often used when transmitting IPs as parameters, such as in the Location header for redirects.
% Thus, despite \engines concealing their \scanips, we can still capture \scanips from various forms of IPs.
\end{itemize}

\noindent\textbf{Sanitized \ipmirrors.} 
Interestingly, \engines are aware that the \ipmirrors can leak their scanning assets, so they mask or replace their scanner IPs.
Specifically, ZoomEye and FOFA substituted the \scanips with placeholders, \ie \textit{``xxx.xxx.xxx.xxx''} and \textit{``*.*.*.*''}, respectively. 
Shodan uses a more advanced method, mapping its \scanips one-to-one to multicast addresses (224.0.0.0/4)~\cite{rfc5771}. Despite these measures, IP addresses can still appear in various protocols and forms, leaving many \scanips visible in search engine results.
%, replacing them in records without leaving any trace, ensuring no real IP addresses are used.

% -----------------------------

% However, it's unfortunate that we discovered that three \engines endeavored to conceal the \scanips within their records. Specifically, ZoomEye and FOFA substituted the \scanips with ``xxx.xxx.xxx.xxx'' and ``*.*.*.*'' respectively. 
% Shodan, which employs a more advanced method, maps its \scanips one-to-one to multicast addresses (224.0.0.0/4)~\cite{rfc5771}, replacing them in records without leaving any trace, ensuring no real IP addresses are used.

\ignore{
\finding{The \engines have taken measures to conceal \scanips in their records, using placeholders (ZoomEye, FOFA) or mapping \scanips to multicast addresses (Shodan).}
}

% attribute/error code

\begin{figure*}[t]
    \centering
    \includegraphics[width=0.8\linewidth]{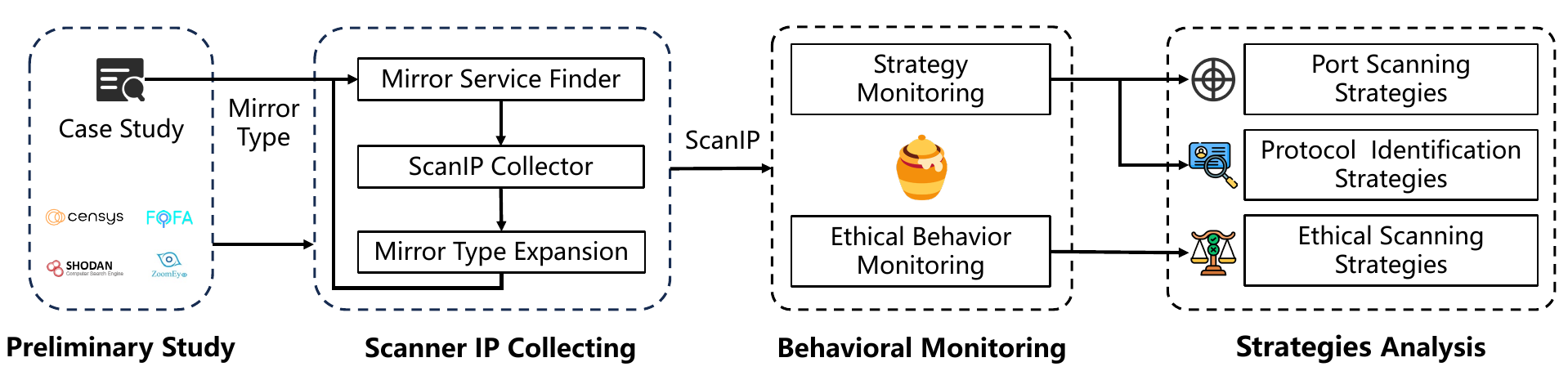}
    \caption{Methodology Overview}
    \label{fig:architecture}
\end{figure*}
\section{Methodology}
\subsection{Overview}
% first challenge
% To understand the behaviors of \engines,\textbf{ the first difficulty lies in differentiating \engine behaviors from those of other bots or potential attackers}, largely due to most search engines' hesitance to disclose their \scanip lists, hindering our tracking efforts. Thus, it is essential 

Based on the preliminary study, we build a two-part framework, including \scanip collection and behavior monitoring modules in this section.
Figure~\ref{fig:architecture} shows the overall architecture of our framework. 
% \del{The \scanip collection module retrieves and filters scanners IP address from \ipmirrors records in \engine results. \wmy{Repeat with the following paragraph}}
% and collects \scanips based on the \ipmirror banners.

\noindent \textbf{Scanner IP Collecting.} To find the scanner IP addresses (\scanip) of \engines,
our preliminary study reveals that the \ipmirror leak visiting IP information in their responses, which can inadvertently reflect the \scanips. Thus, we leverage \ipmirrors to systematically collect \scanips (Section~\ref{sec:ip_collect}).
% Our preliminary study shows that \ipmirror can reflect visit IPs, we can collect \scanips by monitoring \ipmirror records in the \engines.

% another challenge & insight
% As \engine conducts an in-depth analysis of a wide range of Internet protocols and devices, common low-interaction honeypots are insufficient to capture the deep behavior of \engines. 
\noindent \textbf{Search Engine Analyzing.} To systematically understand scanning strategies and capture potential ethical violations by \engines, no off-the-shelf system was available to provide such functionality. Therefore, although it is nearly impossible to simulate every type of device to monitor scanning activities, we instead focused on a highly targeted target: IoT devices, to analyze \engines. The behavior monitoring module utilizes honeypots to collect \scanip behaviors, which trap and collect the behaviors of \scanips based on IoT web honeypots and sniffer honeypots (Section~\ref{sec:behavior}).

% Device search engines are extensively employed for discovering IoT devices, and each \engine offers complicated services for identifying such devices. As a result, we are able to trap deep behavior from \engines effectively.
% elicit? induce? attract? trap?

% The \scanip collection module retrieves \ipmirrors from \engine search results and collects \scanips based on the \ipmirror banners.
% 

\subsection{Scanner IP Collecting}\label{sec:ip_collect}
% 把斜体改掉
% \subsubsection{Architecture}
% Based on our findings in the preliminary study, \ipmirrors, which include incoming IP information in their response echoes, have proven to be consistently effective in discovering \scanips. By analyzing these echoes, we can detect the \scanips used by these engines during their measurements of hosts that employ these services. 

Figure~\ref{fig:ip_collection} in the Appendix illustrates our scanner IP collecting module, which consists of three components: 1) \ipmirror finder for matching records by \ipmirror type patterns; 2) \scanip collector for collecting \scanips from \ipmirror records; 3) Mirror type expansion module for distill new types of mirror service from \scanip records.

% Figure~\ref{fig:ip_collection} illustrates our \scanip collection module, which consists of three components: the \ipmirror finder for match \ipmirrors in \engines by a set of \ipmirror type patterns, the \scanip collector for regularly collecting \scanips from \ipmirror records in \engines, and the \ipmirror type expansion module for expanding new pattern for \ipmirror from \scanip records, which can find new types of \ipmirrors.

\subsubsection{Mirror Service Finder}
% \wmy{recall the goal of each module at the beginning, and summarize the method of this paragraph in one sentence}

% IPMirror Finder模块的目的是批量发现并收集能够泄露IP的服务（IPMirror）。对此，我们提出了一套基于IP泄露pattern数据库(保存了已经收集到的IPMirror的响应特征以及ScanIP匹配模式)，查询测绘引擎上符合特征的记录并进行动静态过滤得到IPMirror的方法
To discover and collect services that can reflect IP (\ie \ipmirror), we propose a methodology that relies on the \ipmirror type pattern. By leveraging the collected \ipmirror type pattern, we efficiently search for relevant records on the engine and subsequently validate the authenticity of the \ipmirror, ensuring accurate and reliable results.

% In our preliminary study, we gathered xx protocols capable of leaking \scanips. Subsequently, we sought to collect \ipmirrors in \engines in bulk. 
% Upon examining the user manuals of these \engines, we discovered that four of them offered a syntax for filtering based on both protocol and response content. \wmy{why mention this here?}

\noindent\textbf{Candidate \ipmirror Collection. }
% Firstly, we combined the service employed by \ipmirror with contextual keywords within its record to establish a pattern denoted as \textit{MirrorPattern} for identifying a specific type of \ipmirror.
% We initiate our approach by extracting search patterns for \ipmirror identified in the preliminary study (Section~\ref{sec:preliminary}).
% % 与ip和环境变量等可变因素无关的文本
% We use the text in the records which are unrelative with the variable factors(IP and environment variables) and the protocol type as its pattern,
% % leveraging protocol type and contextual keywords that don't include \scanip \geng{?} within its records, 
% denoted as \textit{MirrorPattern}.
% %拆成两句完整的话
% % ippattern，mirrorpattern
% For instance, the pattern for \ipmirror reflecting \scanip via the SIP protocol is defined as ``protocol:sip \&\& banner:received=''. Meanwhile, we designate the patterns that match the candidate \scanip in the responses as \textit{IPPattern}. For example, the \textit{IPPattern} corresponding to the SIP protocol is ``received=\$\{ipv4\}''. Then we define the tuple <\textit{MirrorPattern}, \textit{IPPattern}> as a pattern capable of detecting \ipmirrors and mining \scanips.
% % 
% We then convert \textit{MirrorPattern} into the corresponding syntax of the \engine and obtaining matching host records by search APIs. 
We begin collecting candidate \ipmirror instances based on known \ipmirror type identified in Section~\ref{sec:preliminary}. As shown in Figure~\ref{fig:ipmirror}, we first leverage service attributes (\eg \texttt{SIP}) and invariable keywords (\eg \texttt{received}) within records to query and retrieve relevant \ipmirror data. These keywords are manually defined when a new kind of \ipmirror is found and subsequently translated into engine-specific syntax, such as \texttt{protocol="sip" \&\& banner="received"} for FOFA and \texttt{service:"SIP"+banner:"received"} for ZoomEye. Getting the queries, we executed efficient searches through APIs to gather matching host records on each engine.

% During experimentation, we discovered that some host records meet the \textit{MirrorPattern} requirements but harbor fake visiting IP. These records typically originate from honeypot servers that utilize fixed echo content during the simulation of certain services. To filter out such spurious \ipmirrors, we combine static and dynamic verification methods to validate the hosts, ensuring their suitability as observation stations. \wmy{condense it}

\noindent\textbf{\ipmirror Verification}. 
However, not all hosts that meet the pattern requirements are \ipmirrors, as some servers may counterfeit their response (\eg, honeypots). To address this, we propose verifying the records through both static and dynamic methods.
Static filtering is based on two observations: (1) a valid \scanip must be a legitimate public IPv4 address, rather than a private, multicast, or reserved address, and (2) a server typically does not crawl itself using its own IP, meaning the \scanip should differ from the host IP. Dynamic verification involves actively probing the host of the candidate \ipmirror. If the response contains the sender's IP, the server is confirmed as a valid \ipmirror, and the IP in the response record represents the engine's scanner IP, i.e., \scanip.

% Furthermore, we made our web honeypots a special \ipmirror service by responding with the sender IPs, which helped us confirm \scanips more confident.\geng{?}

\subsubsection{ScanIP Collector}
As \engines periodically scan the same host to identify new services and update records, we can acquire \scanips by monitoring the results of \ipmirrors on different \engines periodically.
Leveraging the search API provided by device search engines, we daily query all \ipmirrors from each \engine and extract \scanips.

% Leveraging the search API provided by device search engines, as examples in Table~\ref{tab:engine_syntax}, we daily gather all \ipmirrors from each \engine and extract \scanips using \textit{IPPattern}.

% 测绘引擎提供了基于IP地址和端口号精准过滤的语法，见表x。因此，我们每天在所有测绘引擎中检索一遍所有的IPMirror，并基于p_ip从记录中匹配提取scanIP

\subsubsection{\ipmirror Type Expansion}
% 由于测绘引擎可能会使用同一批IP扫描不同的服务，因此同一扫描IP可能会出现在多种能够泄露IP的网络服务的记录中。基于此发现，我们可以在已经发现的扫描IP的基础上不断发现新的网络服务。
As \engines may use the same \scanips to probe different services, the same \scanip can appear in the records of multiple \ipmirrors. 
By analyzing records containing these \scanips and filtering out known \ipmirrors, we group similar records based on their context and semantics. Through manual inspection of these groups, we identify new types of \ipmirrors and establish their service queries for further exploration and discovery. This approach enables us to expand the scope of \ipmirror from known seeds to new types based on observed patterns in collected data.

\subsection{Behavioral Monitoring} \label{sec:behavior}
% behavior monitoring?

To understand potential attack vectors and study behavioral patterns in network security, researchers frequently deploy honeypots~\cite{honeypot}.
These controlled environments allow for the observation and analysis of \engines, bots, and potential attackers' actions. By monitoring interactions with these honeypots, valuable insights into their behaviors and strategies can be obtained.

To comprehensively understand \engine behaviors from multiple perspectives, we enhanced our honeypot infrastructure with two tailored designs.
% 
% \del{We deployed a diverse set of honeypots to understand \engines' strategies comprehensively.} 
This included a full-port closed honeypot and a popular-port open honeypot to unveil port scanning and protocol identification techniques.
% 
% Additionally, a honeypot with the popular ports open was utilized to uncover protocol identification techniques through captured interactions. 
% 
Further enhancing a web honeypot with IoT device emulation and comprehensive files, we aimed to attract more in-depth scanning sessions to thoroughly evaluate ethical behavior and real-world impact assessments.

\subsubsection{Strategy monitoring}
% \wmy{merge protocol here, move web to 4.3.3}
% strategy monitoring? strategy fingerprinting?
To gain insights into the scanning strategies employed by \engines, we utilized the following two honeypots.

To understand the port scanning strategies, we use a full-port closed honeypot to capture the port scanning activities. Interactions between services on different ports can introduce biases in data packet counts. 
% complicating port scanning behavior analysis.\geng{tbd} 
To ensure equal scanning across all ports, we closed them on the honeypot, allowing each port to receive only one data packet per scan. 
% This approach provides a consistent and unbiased dataset for analysis.
% receive only one data packet per scan? why?

% \subsubsection{Protocol interaction}
% intro
To delve into how \engines discern protocols on open ports, we established a honeypot with commonly used ports open. Due to resource constraints, we concentrated on the top 100 high-traffic ports based on the results from our full-port closed honeypot. Our honeypot passively acknowledged packets without other active responses at the application layer. 
% This approach ensured that only protocol identification traffic from \scanips was captured. To ensure accuracy, non-experimental ports were disabled.
% and essential services were relocated to high-numbered ports to minimize traffic interference.
% passively acknowledged TCP packets? UDP ??

Additionally, we implemented a traffic monitoring function in the honeypots to capture and analyze incoming data packets comprehensively. This enables detailed analysis of scanning patterns and behaviors exhibited by \engines.

\subsubsection{Ethical behavior monitoring} 
As unethical behavior could potentially exist across various services, the exhaustive simulation of all services to capture such behavior is impractical. Thus, we leverage web honeypots as they offer heightened customization and facilitate the emulation of a broad array of web-based services, making them an efficient choice for capturing engines' ethical behavior.

% In addition to analyzing port scanning strategies, we implemented a web honeypot to investigate \engines' web service scanning patterns.
% % 
% The web honeypot hosted an HTTP service to record interactions with \engines.  This enabled us to capture scanning behaviors and strategies aimed at web services.

\noindent\textbf{Customized default pages for IoT devices.}
In the countless web services, we opted to focus on IoT devices, which are abundant in number and often riddled with vulnerabilities. Targeting IoT devices increases the likelihood of capturing anomalous scanning behavior by \engines. IoT devices come in various types with significant differences in functionality. Therefore, we embedded the fingerprints of the IoT device management page into the default homepages of our web honeypots to simulate these devices. To ensure consistent data collection, we configured it to respond to all unknown path access requests uniformly.

% 多加几句unauthentic
\noindent\textbf{Decoy paths.}
To gain insights into whether engines attempt to access sensitive data from hosts without proper authentication and their handling of such content, we constructed a series of decoy paths. Specifically, given that IP cameras represent another common type of IoT device with web services, we referred to various generic configurations of IP cameras and selected 21 typical paths for simulation. These paths can return sensitive information such as camera snapshots and device configuration files. To enhance the authenticity of the simulation, the camera snapshot paths also include dynamic timestamps to simulate real-time monitoring scenarios.

\noindent\textbf{Dynamic trackable links.}
To delve deeper into the scanning behavior boundaries of \scanips, we implemented a dynamic linking strategy within our honeypot. Specifically, we encoded information such as client IP, port, honeypot IP, timestamp, and others for each access and embedded them as clickable links within the page body. This approach introduces variability to the links displayed with each page load, thereby increasing the complexity and uncertainty of the scanning process. 
% Additionally, we simulated common bot scanning paths (such as robots.txt, sitemap.xml, security.txt) and embedded dynamic links within these paths. 
% To distinguish between scanning behaviors, we crafted custom paths with varying depths, enabling differentiation between regular scans and those conducted post-file inspection like robots.txt.
% For the robots.txt file, we deliberately listed 10 paths related to web vulnerabilities in 2023 and pretended that these paths existed but were prohibited from access.

\subsection{Implementation and Result}

Based on our preliminary study of the seven successfully accessed engines, we selected all of those that offer sufficient queries and batch-automatable search API capabilities to collect the Scanner IP. As a result, we focused on four engines: Censys, Shodan, FOFA, and ZoomEye. While our findings may not apply to all engines, our study provides valuable insights into these four search engines, which are widely used in academic research such as \cite{nasr_chargeprint_2023,sasaki_exposed_2022,zhao2020large}, and are likely representative of broader industry trends. The details of implementation can be found in Appendix~\ref{sec:implementation}.

\noindent \textbf{Dataset.}
We deploy our honeypots in 4 cities, including Tokyo, Singapore, Beijing, and Shenzhen. In each city, we deploy five web honeypots, one closed honeypot and one open honeypot, with the same settings, to gain further insights into variations across different regions.
Our data was collected from 28 honeypots deployed across 3 different countries from March 2023 to March 2024. 

\begin{table}[t]
    \centering
    \caption{Overview of \ipmirrors and \scanips detected across four \engines, with data collection from March 2023 to March 2024. The total number of Mirrors is the union of Mirrors from four engines.}
    \label{tab:dataset}
    \resizebox{\linewidth}{!}{
    \begin{tabular}{c c c c}
        \toprule
        \textbf{Engine} & \textbf{\# of \ipmirrors} & \textbf{\# of \scanips} & \textbf{\makecell{\# of \scanips\\ in Honeypot}}\\
        \midrule
        Censys~\cite{censys_search}& 45,580 & 481 & 140 \\
        Shodan~\cite{shodan} & 611 & 91 & 81 \\
        FOFA~\cite{fofa} & 58,671 & 668 & 579 \\
        ZoomEye~\cite{zoomeye} & 3,197 & 167 & 39 \\
        \midrule
        Total & 106,132 & 1,407 & 839 \\
    \bottomrule
    \end{tabular}
    }
\end{table}

In the one-year dataset, in addition to the five seed mirror types in the preliminary study, 74 new mirror types were discovered after mirror type expansion, such as illegal visitor warnings in Redis and the welcome banner of ZXFS FTP. This helped us discover 1183 new \scanips.
As shown in Table~\ref{tab:dataset}, 
%by examining xxx records, 
we found 106,132 \ipmirrors and 1,407 \scanips in the four \engines.
FOFA boasts a significant number of \scanips, totaling up to 665, and maintains the largest number of \ipmirrors. In contrast, the remaining three \engines only possess 91 to 481 \scanips\footnote{While Censys publishes its IP ranges~\cite{censys-ip}, these ranges lack specificity and may introduce false positives in traffic analysis. Therefore, we collected Censys IPs using our own methodology.}. 

Our honeypots captured 7,362,701 requests, with 32,035 in full-port closed honeypots, 347,784 in popular-port open honeypots, and 6,982,882 in web honeypots, from 839 unique \scanips totaling 4.6GB in size. Here we define a request as a transport-layer TCP/UDP packet. Once we have identified these \scanips, we employ the earliest and latest timestamps of their records on \engines to establish their active duration. We then filter and retain only the behaviors exhibited by these \scanips within our honeypot during this designated period, ensuring that the data we collect indeed originates from \engines activities.
% 
% Moreover, we retained results corroborated by many \scanips and communicated with relevant companies regarding individual IPs exhibiting anomalous behavior to avoid inaccurate accusations. \wmy{@hg Will this remind them of our weakness? I'm afraid the reviewer will ask}
%
Due to the different scan strategies across various engines, our honeypot can hardly capture behaviors from all \scanips. Consequently, all subsequent behavioral analysis will be based solely on the subset of \scanips observable within our honeypot. 

\subsection{Discussion}
Currently, \engines are unable to handle special IP formats in responses that they are not aware of when attempting to mask their IPs. We admit that our paper will remind the engines with mirror service in the three IP formats in Section~\ref{sec:preliminary}, but our methodology remains robust. Mirror service can always generate new and diverse methods to encode IPs into the response, such as 1.2.3.4 to 4-3-2-1 or 4\%3\%2\%1, making engines hard/fail to sanitize scanner IPs. 
This will lead to an ongoing iterative battle between device search engines and mirror services.

% \documentclass{standalone}
% \usepackage{subcaption}
% \usepackage{graphicx} % 用于插入图像

% \begin{document}

\section{Scan Strategy}

In this section, we report the scan strategy of \engines according to the dataset acquired from 28 honeypots from March 2023 to March 2024.
% In this and the following sections, we report on our findings on the data collected from 30 honeypots deployed across 3 different countries from March 2023 to March 2024. 

\subsection{Landscape} \label{sec:landscape}

\subsubsection{Geographic distribution}
% \noindent \textbf{Geographic Distribution.}
% {fig/country_without_background.pdf}

\begin{figure}[t]
    \centering
    \includegraphics[width=0.7\linewidth]{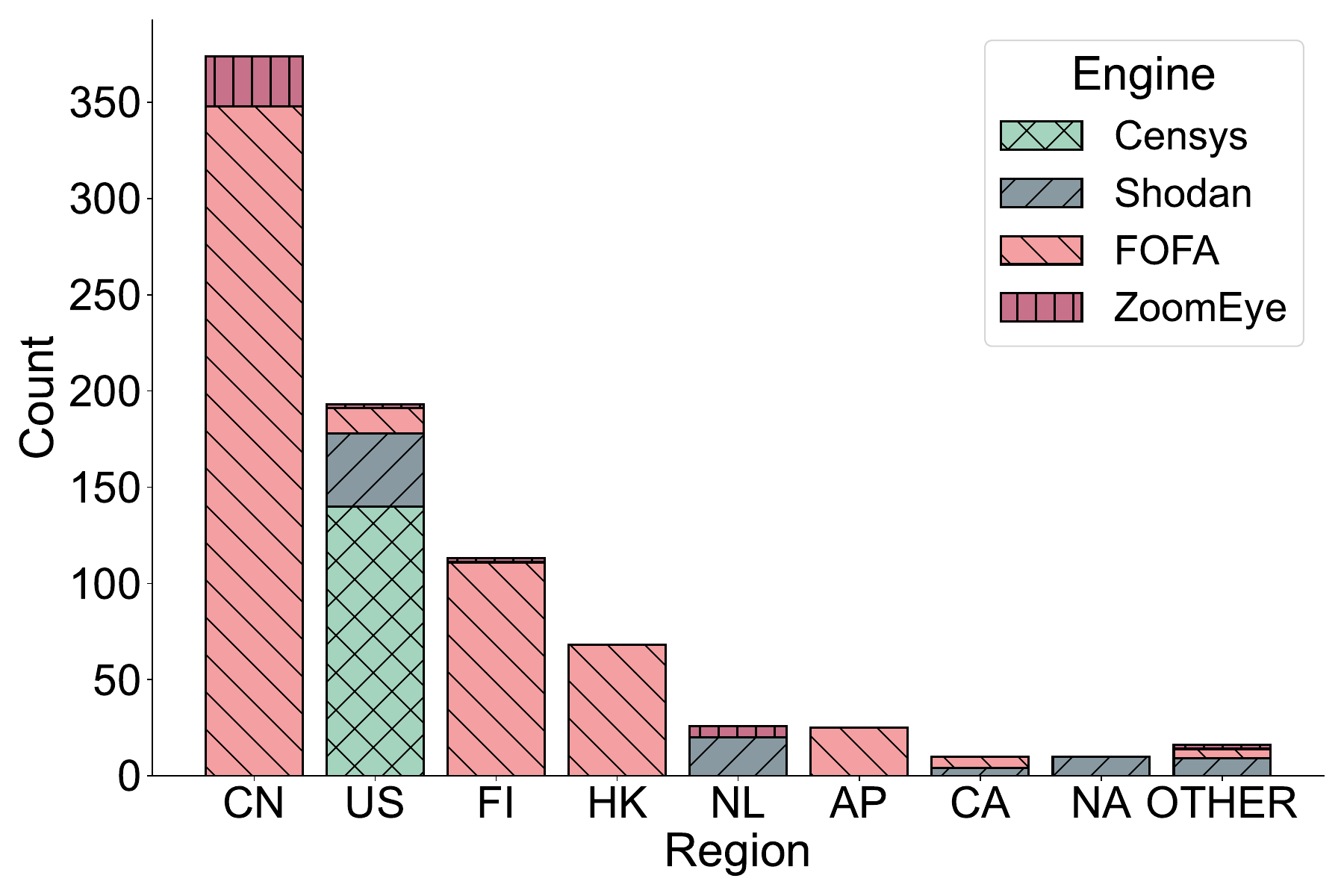}
    \caption{\scanip Region Distribution.}
    \label{fig:region}
\end{figure}
Figure~\ref{fig:region} shows the regions of \scanips used by each \engine. Overall, \engines prefer to use their own country's IPs. For example, 67\% and 72\% of ZoomEye and FOFA \scanips are in China, with FOFA relying on cloud ISPs and ZoomEye using multiple consumer ISPs.  
In the case of Shodan, 46.91\% of its \scanips are from the US, utilizing a combination of both enterprise and cloud ISPs for broader coverage. Censys uses enterprise ISP, brings all its \scanips come from the US.
Additionally, we observed that IPs from Finland and the Netherlands are highly preferred. Specifically, 19.17\% of FOFA IPs and 5.13\% of ZoomEye IPs originate from Finland, while 15.38\% of ZoomEye IPs and 24.69\% of Shodan IPs come from the Netherlands.  
This may arise from their minimal restrictions, competitive prices, and hosting facilities offering high-speed large-bandwidth Internet access~\cite{netherlands_vps}.

\begin{figure*}[t]
    \centering
    \begin{minipage}[t]{0.24\linewidth}
        \centering
        \includegraphics[width=\linewidth]{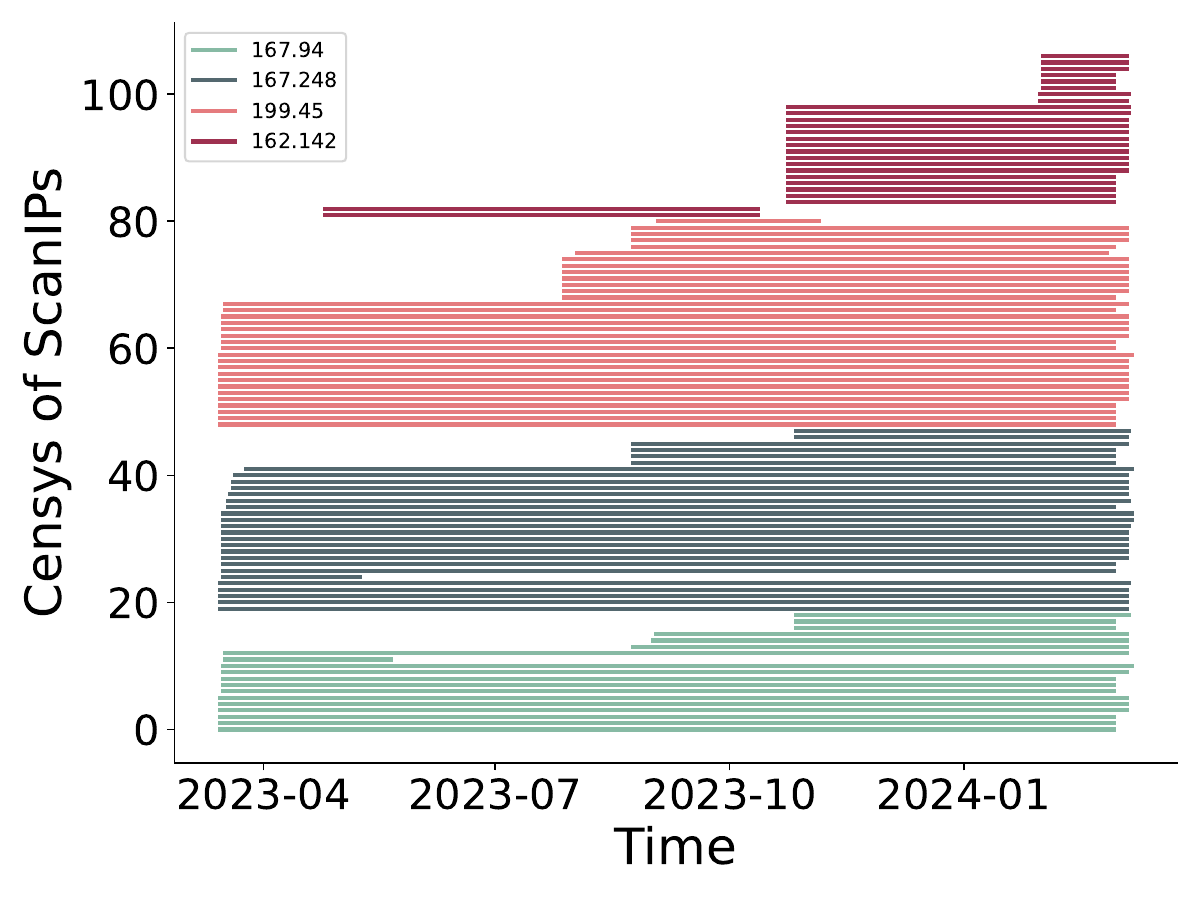}
        \par {\footnotesize(a)} \textnormal{\footnotesize Censys}
        % \label{fig:censys-lifespan}
    \end{minipage}
    \begin{minipage}[t]{0.24\linewidth}
        \centering
        \includegraphics[width=\linewidth]{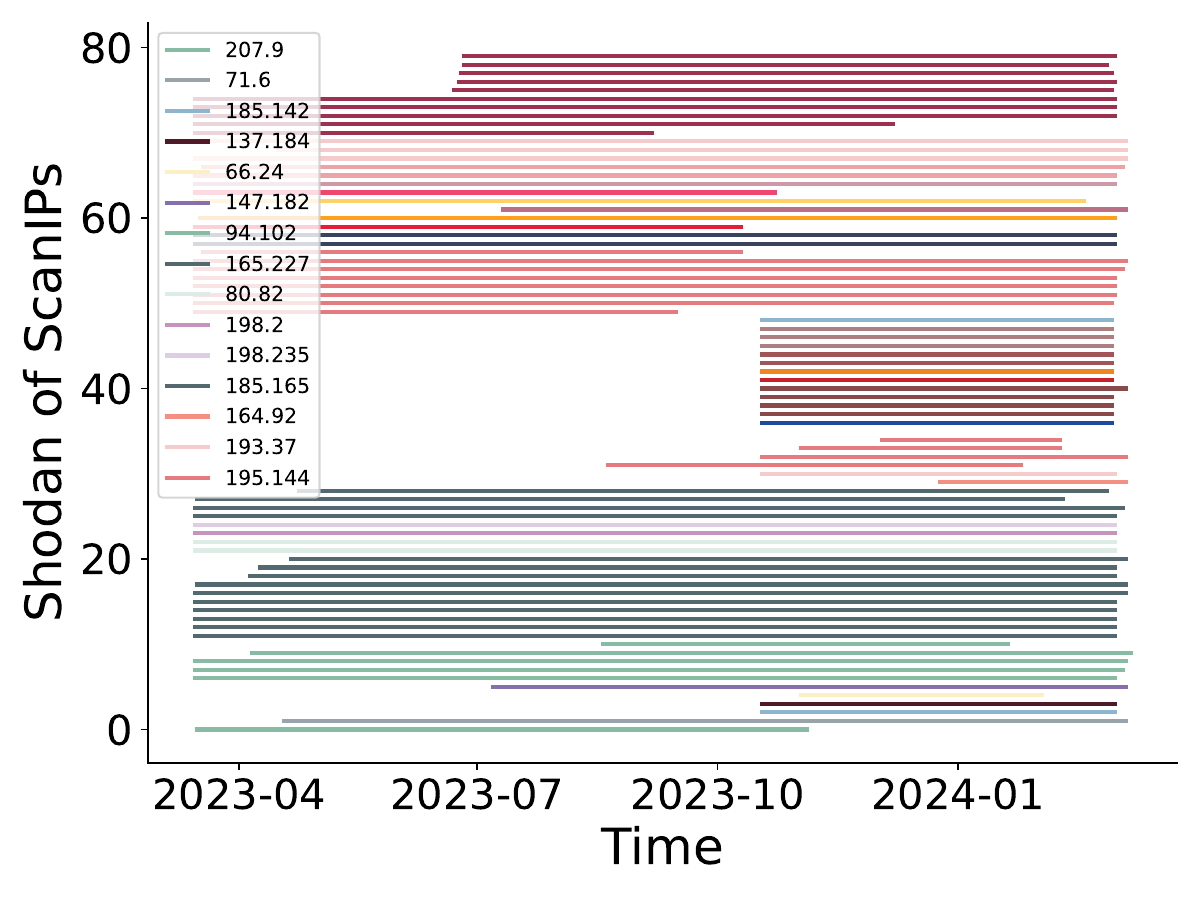}
        \par {\footnotesize(b)} \textnormal{\footnotesize Shodan}
        % \label{fig:shodan-lifespan}
    \end{minipage}
    \begin{minipage}[t]{0.24\linewidth}
        \centering
        \includegraphics[width=\linewidth]{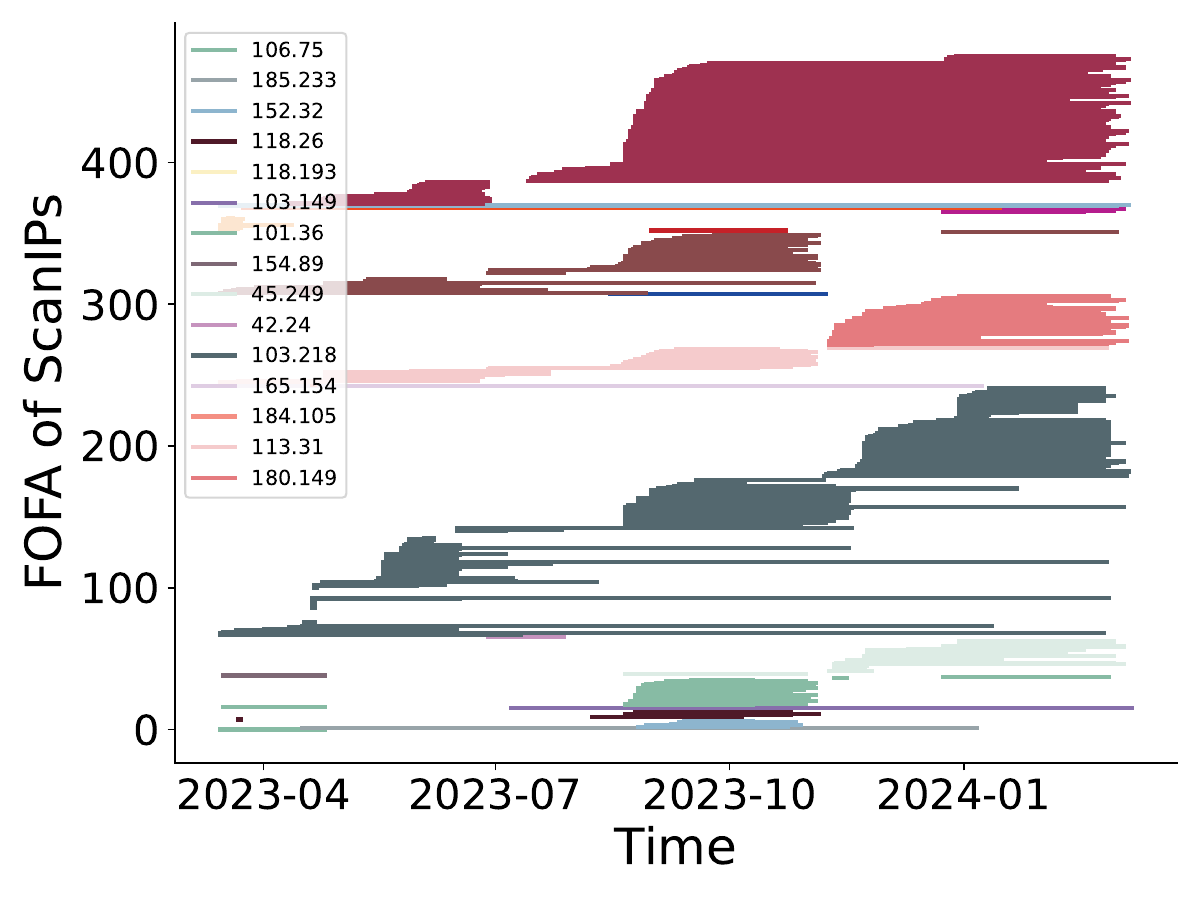}
        \par {\footnotesize(c)} \textnormal{\footnotesize FOFA}
        % \label{fig:fofa-lifespan}
    \end{minipage}
    \begin{minipage}[t]{0.24\linewidth}
        \centering
        \includegraphics[width=\linewidth]{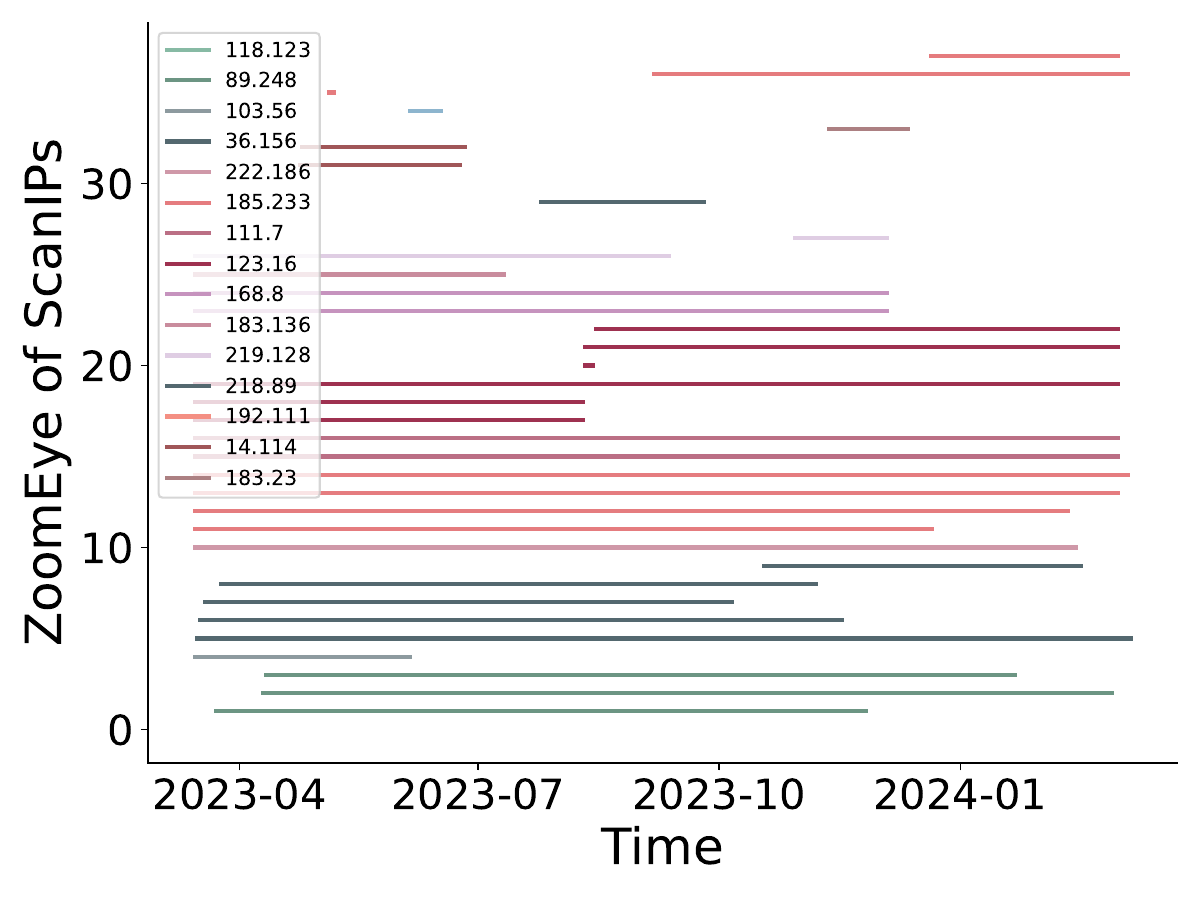}
        % \caption*{ZoomEye}
        \par {\footnotesize(d)} \textnormal{\footnotesize ZoomEye}
        % \label{fig:zoomeye-lifespan}
    \end{minipage}
    \caption{The lifespan of \scanips in Censys and FOFA. Each line represents the lifespan of one \scanip, and \scanips in the same network segment (with a /16 subnet mask) are marked with the same color.}
    \label{fig:lifespan}
\end{figure*}

\subsubsection{Server rotation strategy} \label{sec:ip_usage}
The \scanips usage duration in \engines reveals insights into their operational strategies. We introduce \scanips lifespan as a metric, capturing the time between their first and last appearance in honeypots. Figure~\ref{fig:lifespan} shows the lifespan of \scanips. We observed the lifespan of different scanners overlapping targeting the same mirror service, which indicates that one mirror service is scanned by IPs randomly selected from the \scanips pool.

All \engines engage in the bulk activation of \scanips. For instance, Censys activated 7, 9, and 16 scanners on July 28th, August 24th, and October 24th, 2023, respectively. Similarly, we observed Shodan activate  18 \scanips across 11 network segments on October 20th.

Both ZoomEye and FOFA demonstrate patterns of IP abandonment and rotation. 
In FOFA, a consistent pattern emerges, characterized by four instances of mass \scanip activation: occurring in mid-May, mid-August, and mid-November 2023, as well as early January 2024, followed by their abandonment around the same periods.
% What's worse, ZoomEye informed us that their IPs are dynamically assigned by ISPs, meaning scanner IPs may change at any moment. 
In contrast, ZoomEye's IP changes lack periodicity, as shown in Figure~\ref{fig:lifespan}d. Notably, in our communications, ZoomEye informed us that their IPs are dynamically assigned by ISPs, indicating that their \scanips are not fixed and may change at any moment, which aligns with the lack of periodicity.
In our one-year monitoring, we did not observe Censys and Shodan significantly retiring \scanips.
% 
% We investigated the cloud service FOFA use and found a three-month duration package, which aligns with the observed cycles of \scanips rotation. 

We further checked AbuseIPDB~\cite{abuseipdb}, a blocklist where users report malicious IPs, and found 665 \scanips have been labeled with ``Port Scan'', ``Hacking'' and ``Brute-Force'' tags.
Rotating \scanips makes scanning activities more resilient against being blocklisted
by IPs. 
\ignore{
However, this practice also poses risks. We found three IPs used to belong to FOFA are labeled in malicious scanning activities. 
% Upon communication with FOFA, it was confirmed that these malicious scanning incidents occurred outside of the time periods when they held those IPs. 
This highlights the potential reputational risks faced by FOFA due to the frequent rotation of its IPs.
%, as malicious actors may exploit these IPs for malicious purposes.
We confirmed with FOFA that these IPs were once theirs but are no longer in use by them. }

% cannot be held stable over time and
% 加数字
\finding{FOFA and ZoomEye do not use fixed scanning assets, with FOFA typically rotating its IPs every three months, making it hard for users to avoid being scanned by blocklisting \engine IPs.}
% 难以blocklist

\ignore{
\subsubsection{Regional tendencies} \wmy{remove?} \geng{this section can be condensed to one or two sentences, or just removed since we only have 30 honeypots, such difference may be because of statistical error.}
To assess potential differences in scanning behaviors among various mapping engines across different regions, we deployed honeypots in four distinct cities. Our analysis revealed that the scanning paths employed by these engines did not vary significantly based on geographical location.
However, a notable trend emerged in terms of the quantity of data packets. Specifically, Shodan, FOFA, and Censys all exhibited a higher frequency of scanning activities towards Tokyo compared to the other regions. The average scanning traffic received in Tokyo was 1.28, 2.04, and 1.33 times greater than that in Beijing and Shenzhen, respectively.
This disparity can be attributed to the \engines' scanning of a larger number of ports in Tokyo. For instance, Censys scanned 99 ports in Tokyo, whereas it only scanned 74 ports in Shenzhen. Similarly, FOFA scanned 1,443 ports in Tokyo, compared to 1,063 ports in Shenzhen.
}
%Furthermore, Censys demonstrated a shorter scanning interval in Tokyo. For instance, within the same time frame, IP 167.94.145.57 scanned port 5985 six times in Tokyo, whereas it only scanned four times in other regions.

\subsection{Port Scanning Strategy} \label{sec:port_scan}
% 先说设置不同
% 再说端口偏好
% 列一个表说明扫描的端口不一样
Port scanning is a crucial function of \engines, 
allowing users to identify open ports and their associated services. Although modern scanning tools can efficiently scan IPv4 addresses, due to resource constraints, we find that \engines do not scan all ports of the entire IPv4 space once a day, and make trade-offs between different ports. Based on our port-closing honeypot, we can analyze the scanning preferences of different \engines.

We first examined the packet setting of their port scanning. The \engines use different TCP settings when scanning. 
TTL (Time to Live) indicates the packet's lifespan in the network. ZoomEye stood out with SYN packets having TTL values approaching 240, significantly higher than Shodan(110), Censys(50), and FOFA(50), which is also higher than the default TTL values of Linux/MacOS (64) and Windows (128). 
While a higher TTL increased the probability of packets reaching their destination, it also burdened routers, potentially leading to waste in scenarios with poor network conditions or faults, especially when there are routing loops in the network.
As for TCP window size, 
Shodan dynamically adjusts its size between 1,024 and 65,535, while others use fixed sizes, including FOFA(1,400), Censys(42,340), and ZoomEye(63,540).
A large window can facilitate faster data transmission in unstable network environments. However, considering that engines need to continuously send scanning packets, 
a large window may increase network load.

\begin{table}[t]
    \centering
    \caption{Top 10 ports scanned by each \engines and all visitors except \engines.}
    \label{tab:top_10_ports}
    \begin{tabular}{c c c c c c}
    \toprule
        \multirow{2.5}{*}{\textbf{Rank}} & \multicolumn{4}{c}{\textbf{Device Search Engine}} & \multirow{2.5}{*}{\textbf{Others}} \\
         \cmidrule(l){2-5}
        & \textbf{Censys} & \textbf{Shodan} & \textbf{FOFA} & \textbf{ZoomEye} &  \\
    \midrule
        1 & 443 & 443 & 443 & 443 &23 \\
        \rowcolor{gray!20}
        2 & 3306& 2222& 22& 2222&3389 \\
        3 & 22& 22& 23& 500&445 \\ 
        \rowcolor{gray!20}
        4 & 23& 23& 3306& 53&22 \\
        5 & 2222& 3306& 2222& 161&80 \\
        \rowcolor{gray!20}
        6 & 139& 3389& 123& 5683&6379 \\
        7 & 32080& 53& 53& 9001&443 \\
        \rowcolor{gray!20}
        8 & 43080& 19& 21& 587&8088 \\
        9 & 21& 161& 8443& 5060&8080 \\
        \rowcolor{gray!20}
        10 & 2323& 2087& 5060& 123&1433 \\
    \bottomrule
    \end{tabular}
\end{table}

\noindent\textbf{Scanning range.} 
Despite \engines scanning all 65,536 ports extensively, only the frequently scanned ports represent their interest. We found significant variation in the number of ports targeted by different engines. For example, 20\% of Shodan and ZoomEye's traffic targets 29 ports, whereas Censys scans 49 ports, and FOFA targets seven ports. 
% \geng{what's this sentence mean?}\wmy{give a reason of top 30, prove that they have preference meanwhile}
% Therefore, we focus only on the differences in their top 30 ports. 

We compare the top 10 ports they scanned and all other visitors (excluding the device search engines) in Table~\ref{tab:top_10_ports} and found that \engines have unique preferences compared to the nature, usually real attackers. Ports like 445(SMB), 80/8088/8080(HTTP), 6379(Redis) and 1433(SQL server) are commonly targeted outside the \engines.
Notably, port 443 attracts the most attention from all four engines, as it is the standard port for HTTPS, which has the most possibility to catch web services.
%
% In contrast, port 80 does not attract the same amount of traffic in top 10, the highest rank is 29 in Shodan.
%
Among common HTTP/HTTPS ports, aside from 443, ZoomEye considers port 9001, FOFA and Censys consider port 8443, and Shodan includes ports 10001 and 8009.
As for the other protocols, Shodan, Censys, and FOFA focus on the common services exposed to the public network. The
prioritized scanning ports include SSH (22/2222), Telnet (23/2323), MySQL (3306), and NTP (123).
% \wmy{如果我们放宽一点看top30}
However, ZoomEye's scanning focuses on high-risk targets. When looking at a wide range of scanning ports, 14 of the ports frequently scanned by ZoomEye ports are not preferred as the top 30 by other engines, such as CoAP (5683), game server (27015), and BitTorrent (6881). These services are frequently abused for reflective amplification DDoS attack~\cite{coap_attack, el2007bottorrent}, 
highlighting ZoomEye's unique scanning behavior. This focus on less commonly monitored ports provides valuable insights into potential vulnerabilities and emerging DDoS threats.

\finding{ZoomEye prefers to scan ports with a high risk of DDoS attack, while other engines focus on the most common ports on the internet.}

\subsection{Protocol Identification Strategy} \label{sec:service}
Details of running protocols on network assets are critical threat intelligence, as they help pinpoint potential vulnerabilities and targets for attacks. However, the methods employed by \engines to effectively identify these protocols remain unclear. This study examines how \engine identifies protocol services on the host.
% interacts with these services and shows preferences for different services. \geng{?}
To solve this problem, we first developed a two-step method to identify the probes used by \engines and subsequently conducted an analysis of their probing strategies.

\subsubsection{Methodology}
\ignore{
To discern services operating on ports, \engines deploy targeted service probes, thereafter inferring the active services and protocols from the responses of the port. By analyzing the probes utilized by \engines, we can comprehend how \engines identify services.

However, it is important to note that not all packets carry application layer protocol data, also referred to as valid payload. 
In TCP\cite{tcp_psh} design, the setting of PSH flag in TCP header indicates that receiver should push the data to the application layer immediately.
Therefore, for TCP packets, we exclusively treat the payload of PSH packets as probes. Conversely, UDP packets, being connectionless, do not establish connections like TCP. As a result, every UDP packet's payload serves as a probe. 

% Accurately identifying probes for corresponding protocols is challenging due to the need to distinguish such packets from a vast amount of traffic data. \geng{why? plz clearly tell the reason} 
% Probe identification is challenging due to the lack of a general parser, which can identify specific application layer protocols from a large number of different protocol packets. 
}

Identifying protocol-specific probes is not straightforward due to two challenges: (1) the probes sent by \engines are intended to identify a wide range of protocols, leading to diverse probes with different payloads in the traffic, and (2) even within the same protocol, variations occur due to different versions or configurations, resulting in inconsistent probes of one protocol.

% 难点在于缺乏一个通用的解析器，能够从大量不同协议的报文中识别出具体的应用层协议以及功能。

\noindent\textbf{Rule generation.} We developed a comprehensive set of rules that can encompass a broader range of protocols by using the existing common rule list and manually adding more rules.

\begin{figure}[t]
    \centering
    \includegraphics[width=\linewidth]{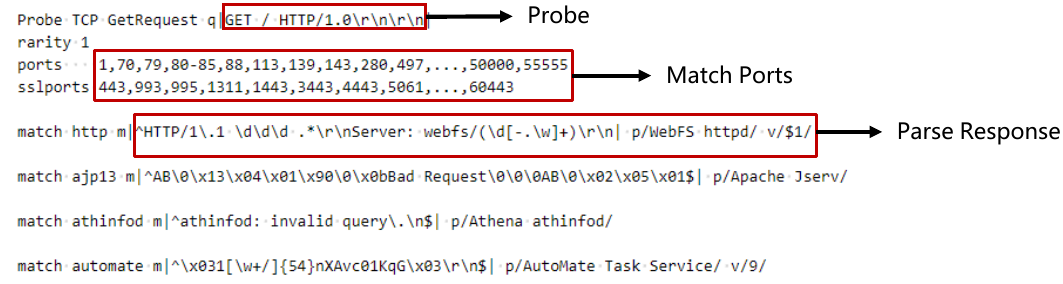}
    \caption{A typical probe rule in nmap-service-probe~\cite{nmap-service-probes}, including service probe, match ports, and the response to parse. We utilize the probes in this list to identify part of the payloads from \engines. }
    \label{fig:nmap_probe_service}
\end{figure}

Firstly, we utilized the rule list from nmap-service-probes\cite{nmap-service-probes}, which contains probes for querying various services and matching expressions to recognize and parse responses. Figure~\ref{fig:nmap_probe_service} shows an example of ``GetRequest'' probe identifying an HTTP service.
% We employed regular expressions to extract 100 types of TCP probes and 84 types of UDP probes from this rule list. Subsequently, using these rules to match payloads in the packets allowed us to identify 25 types of TCP probes and 31 types of UDP probes sent by \engines.
% 
% Initially, when dealing with prevalent protocols exhibiting numerous variants, 
Besides, we employ existing network package parsers to identify other services, such as the Scapy library~\cite{Scapy} for the TLS protocol.

% We utilize the TLS parsing function provided by the Scapy library~\cite{Scapy} to process payloads, thereby improving the recognition rate of TLS payloads. 

% This approach may result in missed detections because Nmap cannot cover all payload types gathered by each \engine, and mismatches can occur due to optional attribute fields in most protocols.

% 有variants， 可能是版本或者option导致的变化
% 1. cover 更多protocol
% 2. 更generic

\noindent\textbf{Fuzzy matching.}
To handle the various probe variants, we refined the matching method using domain-specific protocol knowledge, enhancing the generality of the matching process.
%
% To address the issue of matching failure caused by optional attributes\geng{sorry for not understanding this goal, what's the relationship between this paragraph with the previous}, 
% 
% based on the empirical observation that payloads of the same protocol generally have similar formats, we optimize the matching method by using existing parsers and regular expressions. 
% 参考某些协议的？编写通用的正则规则去匹配payload
% Additionally, based on our domain knowledge of some protocols, we improved the existing matching rules using regular expressions. 
% \wmy{利用编辑距离小的去猜是什么协议}
We first calculate the edit distance between unmatched probes and the acquired rule list. Then, we select the rule with the smallest distance and determine if the corresponding protocol is related to the variant.
For instance, the probe for Oracle TNS in nmap rule list is ``\texttt{\textbackslash x00Z\textbackslash x00\textbackslash x00\textbackslash x01\textbackslash x00\textbackslash x00\textbackslash x00...}''. However, according to the design of Oracle TNS\cite{oracle_tns_format}, the first two bytes (\ie \texttt{\textbackslash x00Z}) indicate the packet length, which is a dynamic value across packets. We improved it by ``\texttt{\textbackslash x00*\textbackslash x00\textbackslash x00\textbackslash x01\textbackslash x00\textbackslash x00\textbackslash x00...}'', where \texttt{*} represents a wildcard. 

Since off-the-shelf rule lists cannot cover all probes, we also manually survey the remaining unmatched probes. Specifically, we searched for unmatched payloads in the form of hexadecimal escape characters on Google, then inferred the purpose of the probe based on the query results.

\subsubsection{Results}
% 最终，我们利用前面的规则在四个测绘引擎上识别出98种tcp payload和86种udp payload，如表6.x所示。
% We collected traffic data for one month from March 8, 2024, to April 7, 2024. 

\begin{figure}[t]
    \centering
    \includegraphics[width=\linewidth]{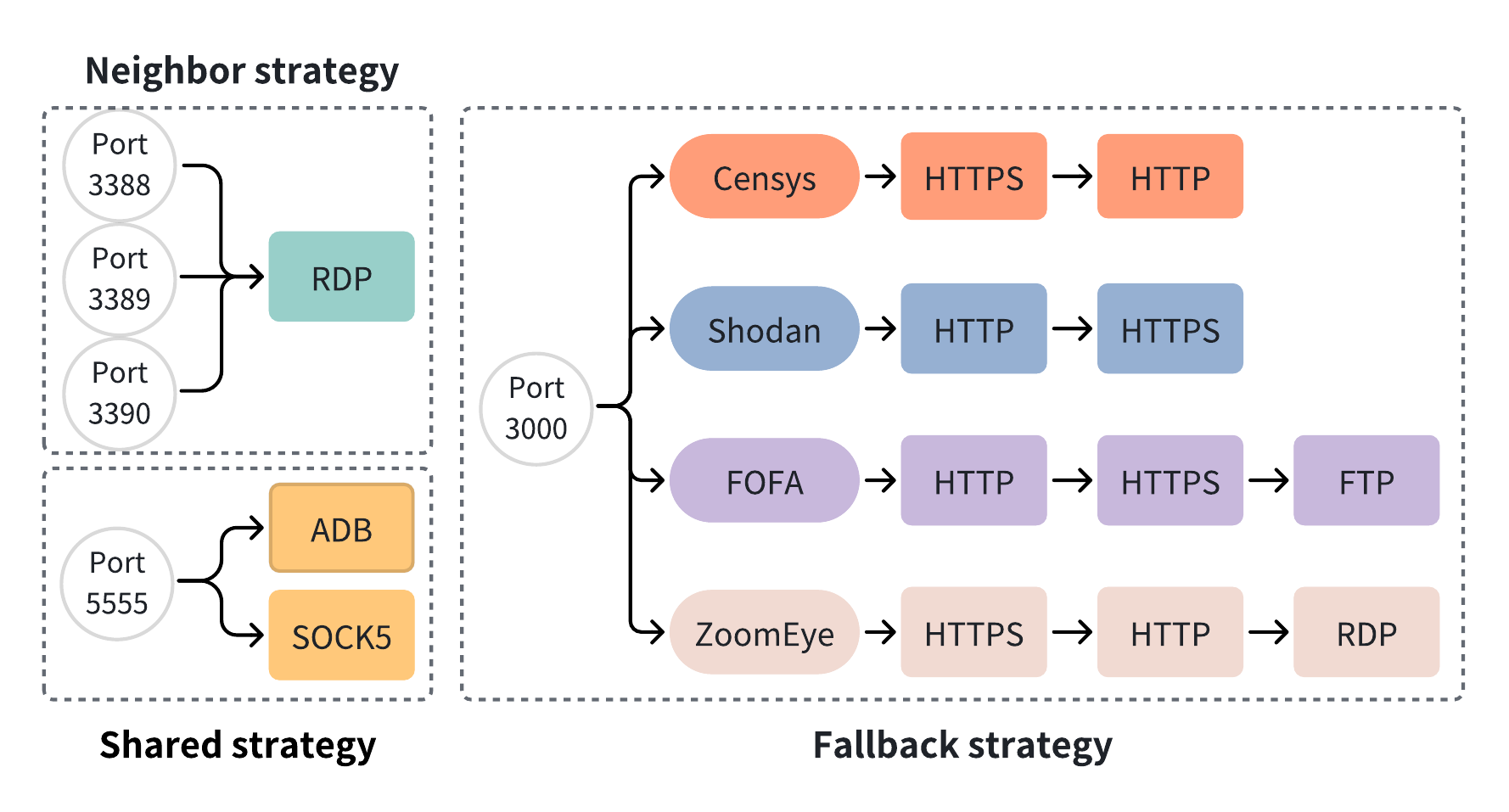}
    \caption{Three probe strategies across the four engines. The order of fallback probes is sorted according to the sequence of probes.}
    \label{fig:probeStrategy}
\end{figure}

% 77,099 个有效的packet中识别出73153个
Analyzing the traffic captured by our popular-port open honeypots, we identified 60 types of TCP probes and 67 types of UDP probes targeting 42 protocols, covering 94.8\% of the packets, as illustrated in Table~\ref{tab:probe_port}. We summarize three different strategies as shown in Figure~\ref{fig:probeStrategy}.
% 出现次数、匹配成功的次数、来自那些规则
% two types of probes with
\begin{table}[t]
    \centering
    \caption{The number of protocols and ports of the identified probes in different \engines.}
    \label{tab:probe_port}
    \resizebox{\linewidth}{!}{
    \begin{tabular}{c c c c c}
        \toprule
        \textbf{Engine} & \textbf{\makecell{\# of \\TCP\_Protocol}} & \textbf{\makecell{\# of \\TCP\_Port}} & \textbf{\makecell{\# of \\ UDP\_Protocol}} & \textbf{\makecell{\# of \\ UDP\_Port}} \\ 
        \midrule
        Censys & 20 & 72 & 23 & 372 \\ 
        Shodan & 21 & 80 & 34 & 51 \\
        FOFA & 26 & 72 & 6 & 6 \\ 
        ZoomEye & 16 & 52 & 32 & 118 \\  
        \bottomrule
    \end{tabular}
    }

\end{table}

% Censys & 20 & 72 & 23 & 372 \\ 
% Shodan & 21 & 80 & 34 & 51 \\
% FOFA & 26 & 72 & 6 & 6 \\ 
% ZoomEye & 16 & 52 & 32 & 118 \\ 

% \wmy{Give each paragraph a subheading, and change into plaintext}\song{Done!}
% Specific Probe是指针对某一类协议生效的探针，一般是特定协议的握手报文。例如，TLS_handshake,X11Probe等。这种探针常用于探测特定协议对应的默认端口号，一般只有运行特定协议的服务器才会响应此探针，从而此类探针识别结果的准确率更高。
% 与之相对的是Generic Probe，这类探针能够对多种不同的协议生效，原因是这些协议在设计的时候都实现了对该探针的响应支持。这种探针可能会出现在多种不同协议对应的默认端口号上，通过建立不同协议的响应匹配规则来识别运行的协议信息，识别准确率会有所下降。
\noindent\textbf{Probe Types.} The probes we collected can be classified into two categories based on their corresponding protocol: \textit{Specific Probe} and \textit{Generic Probe}.

\textit{Specific Probes} are effective for a specific protocol, generally the handshake messages of a particular protocol, such as ``\texttt{\textbackslash x6C\textbackslash x00\textbackslash x0B\textbackslash x00\textbackslash x00\textbackslash x00...}'', which is a hello message of X11~\cite{X11_rfc}. These probes are primarily used to detect default port numbers associated with specific protocols.
% , resulting in higher accuracy in identification.

% 'RPCCheck','Help','GenericLines' in Nmap @wmy
On the other hand, we found three \textit{Generic Probes} that are designed to be effective across multiple protocols, which share the same command or handshake method. For instance, the probe ``help\textbackslash r\textbackslash n'' is applicable to various services, including ident\cite{ident_rfc}, SMTP\cite{smtp_rfc}, NNTP\cite{nntp_rfc} and so on. 
% \geng{why do we mention this sentence?} The probes are all Nmap's probes, including ``RPCCheck'',``help'' and ``GenericLines'', and they may be encountered on default port numbers associated with different protocols.
%, potentially reducing recognition accuracy when establishing response matching rules to identify protocol information.

% 添加help指令的解读
% 多种协议共享了指令 支持同样的handshaking方式
% 有限状态机的图

% \begin{itemize}
%     \item \textbf{Specific Probe}: Tailored to a particular protocol, such as TLS\_Handshake, X11Probe, etc. 
%     \item \textbf{Generic Probe}: Not specific to a particular protocol but capable of detecting multiple service types. For instance, the probe ``\textbackslash r\textbackslash n\textbackslash r\textbackslash n'' is applicable to various services, including FTP, ident, POP3, UUCP, Postgres, and whois\cite{nmap_generic_probe}.\wmy{add citation}
% \end{itemize}

\begin{figure*}[t]
    \centering
    \includegraphics[width=\linewidth]{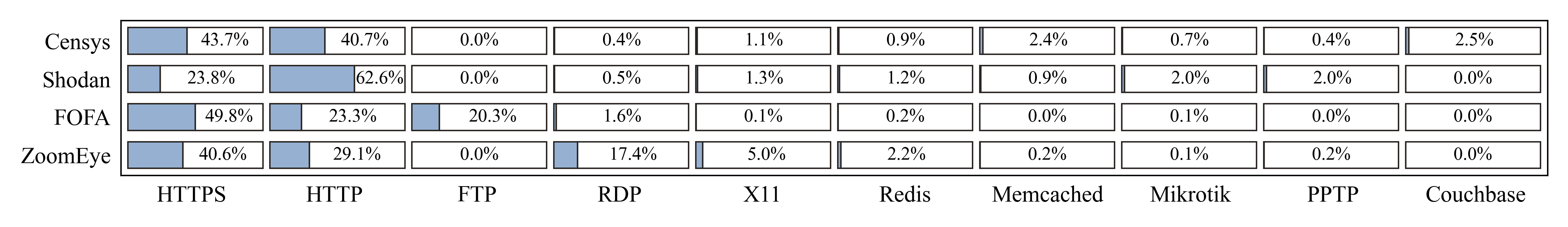}
    \caption{Top 10 protocol/services that have the highest proportions across the four engines.}
    \label{fig:top10_service}
\end{figure*}

% We discovered:

\noindent\textbf{Neighbor strategy.} 
Table~\ref{tab:service_multiport} in the Appendix shows the identified protocol payloads alongside their corresponding port numbers.
% 
% 从探针使用顺序上可以发现，测绘引擎会优先使用特定探针去探测那些与协议存在默认对应关系的端口列表，
% 总的来看，使用特定探针的种类越多，测绘引擎的协议覆盖率越高，其提供的测绘资产就会更加全面。
% Device search engines employ probes that are specific to certain protocols. During the detection of port protocol information, these \engines typically choose probes that correspond to the default protocol. 
There is no doubt that \engines prioritize utilizing Specific Probe to identify port services associated with default services, such as requesting DNS\cite{rfc1035} probes on port 53. 
Additionally, beyond default service ports, \engines also attempt to probe services on certain neighbor ports. For instance, although the default port for the X11 protocol~\cite{X11_rfc} is 6000, we observed X11 probes being received on ports ranging from 6000 to 6002. Similarly, we observe RDP~\cite{rdp_protocol} being probed on ports 3388 to 3390, despite 3389 being the default port. Neighbor ports also include jumping ones, such as 5673 VS 5683(CoAP), and 6666/7000 VS 6379(Redis).
Service deployers who wish to avoid identification should refrain from using default ports of protocols, as well as neighbor ports we listed in Table~\ref{tab:service_multiport}.

% \wmy{有人把端口迁移想规避，这有效吗 改成RQ2}
\finding{Users cannot evade scans by migrating the ports of services they wish to hide because \engines probe protocols not only on default ports but also on neighbor ports.}

\noindent\textbf{Shared strategy.}
Some ports are used by multiple protocols, instead of one specific protocol, leading to potential collisions. Therefore, multiple probes from various potential protocols are sent to the same port.
For example, probes for both the adb~\cite{adbconnect}(Android Debug Bridge) and socks5~\cite{socks5} protocols were received on TCP port 5555. 
% These ports generally do not have a strong association with a specific protocol but are shared by many different services or protocols, thereby collecting probes from various potential protocols. \geng{tbd}

\noindent\textbf{Fallback strategy.} When \engines fail to identify the protocol on specific ports, they employ a fallback strategy to explore alternative protocols, as shown in Figure~\ref{fig:probeStrategy}. Consequently, multiple probes are observed across a majority of ports. 
% 
% 此类端口一般未与某类协议存在强绑定的关系，而是被许多不同的服务或协议共享，从而会收集到来自多种潜在不同协议的探针。
% \finding{To improve the protocol identification rate, {\engines} also attempt to use probes for multiple different protocols on the same port.}
% \ww{The \engines also have a more general strategy, 
All four \engines employ a combination of \texttt{GET HTTP} and \texttt{TLS} handshake to enhance web service detection. Moreover, FOFA and ZoomEye have incorporated FTP and RDP probes into their fallback strategies, respectively. 

% Moreover, we have identified several intriguing findings. Beyond the default ports of protocols, certain ports, which are frequently utilized in practice, also contribute to the domain knowledge for the engines. For instance, the default port for the X11 service is 6000, but we observed that X11 probes from the engines have been received on ports ranging from 6000 to 6002. A key takeaway here is that service deployers who wish to avoid being mapped should not only steer clear of the default ports of protocols but also avoid neighboring or commonly used ports.

% We have also discerned that some ports have accumulated payloads for multiple protocols. For instance, probes for both the adbConnect and sock5 protocols were received on TCP port 5555. This leads us to another important finding: to enhance the rate of protocol identification, the engines also endeavor to deploy probes for multiple distinct protocols on the same port.

% 此外，我们还有一些其他的有趣发现
% Furthermore, we have some other interesting findings.

\noindent\textbf{Protocol preference.}
According to the probe's aim to protocol, we can learn the different preferences of \engines in identifying protocols. Figure~\ref{fig:top10_service} shows the top 10 protocols/services that have the highest proportions across the four engines.
Although 443 is the favorite port of all engines, Shodan is trying to find more HTTP(62.6\%) services, compared to HTTPS(23.8\%). 
FTP for FOFA(20.3\%) and RDP for ZoomEye(17.4\%) stand out, matching the specific generic fallback strategy unique to each engine while having a much smaller presence in the other engines.
% Other protocols like Redis, Memcached, Mikrotik, PPTP, and Couchbase show lower, but varying levels of scanning activity across the engines, indicating some level of interest in these services as well.

\section{Ethical Scanning} \label{sec:ethics}
In this section, we critically evaluate the ethical practices of \engines from three aspects: transparency, harmlessness, and anonymity.

The operations of these engines involve accessing computer systems and collecting sensitive data, raising important ethical considerations. To safeguard citizens' computers and data, various countries have enacted robust cybersecurity and personal information privacy laws, such as the European Union’s GDPR~\cite{gdpr} and NIS2~\cite{NIS2}, the USA’s CFAA~\cite{CFAA} and CCPA~\cite{CCPA}, as well as regulations in China~\cite{ChinaDataSecurityLaw,ChinaCyberSecurityLaw,ChinaPersonalInformationProtectionLaw}, Japan~\cite{JapanUnauthorizedComputerAccessLaw,JapanPersonalInformationProtectionLaw}, and Singapore~\cite{SingaporePersonalDataProtectionAct,SingaporeComputerMisuseAndCybersecurityAct}.
They apply to the countries\footnote{The inclusion of European Union regulations is essential due to the international nature of data flow and the stringent requirements of GDPR.
Article 3, paragraph 2(b) of the GDPR~\cite{gdpr} stipulates that
%: This Regulation applies to the processing of personal data of data subjects who are in the Union by a controller or processor not established in the Union, where the processing activities are related to the monitoring of their behavior as far as their behavior takes place within the Union.}
the regulation applies to any entity processing and monitoring the data of EU citizens, regardless of the entity's location.}
where the \engine companies are registered and where our honeypots are deployed. 

Although there are currently no specific legal interpretations or industry standards explicitly applicable to \engines, we propose a set of ethical principles aimed at safeguarding users' rights. These principles draw from best practices established by notable tools and engines such as ZMap, Censys, and Onyphe~\cite{durumeric_zmap_2013,durumeric2015search,onyphe-standard}, guidelines for search engine crawlers~\cite{rfc9309,guideline_robot}, and foundational ethical frameworks like the Menlo Report\cite{dittrich2012menlo}. Key practices include transparency, harmlessness, and anonymity. Our evaluation results in these three areas are summarized in Table~\ref{tab:ethical violation}.

\ignore{
\begin{table}[ht]
    \centering
    \begin{tabular}{c|c}
        Country & Law \\
        EU & General Data Protection Regulation \\
        US & California Consumer Privacy Act \\
        China & Personal Information Protection Law \\
    \end{tabular}
    \caption{Laws about cybersecurity and information privacy. The regions involve the }
    \label{tab:my_label}
\end{table}
}
% informing users, conducting non-invasive scanning, minimizing data collection, and showing anonymized data, 
% Then check the actions of the \engines.
% \footnote{Due to unsuccessful registration attempts and several unresponsive inquiries to Onyphe, this study did not include Onyphe in its research scope.}

 % from four aspects: whether they inform users during scanning, whether their scanning of assets is intrusive, and whether they over-collect device privacy

\ignore{
Device search engines are not merely concerned with technological accuracy; they also uphold the utmost respect and protection for personal privacy, information security, and even social order and customs.
Despite their convenience in providing critical data, these engines may potentially impact hosts and networks across the internet.
As such, it is imperative to conduct thorough investigations into the ethical performance of various \engines during their scanning activities. 

We evaluated the ethical violation of \engines from the three aspects: right to be informed, intrusive scanning, and privacy leakage, to assess the potential risks. We summarized our results of this section in Table~\ref{tab:ethical violation}.
}

\begin{table*}[h]
    \centering
    \caption{Ethical violation of \engines.}
    \label{tab:ethical violation}
    % \resizebox{0.9\linewidth}{!}{
    \begin{threeparttable}
    
    \begin{tabular}{c l l c c c c}
        \toprule
        \textbf{Type} &  \multicolumn{2}{l}{\textbf{Action}} & \textbf{Censys} & \textbf{Shodan} & \textbf{FOFA} & \textbf{ZoomEye} \\ 
        \midrule
        % \rowcolor[20]{grey}
        
         & \multicolumn{2}{l}{Explain the purpose on every probe.}  & \neutral & \evil & \evil  & \evil \\ 
        \rowcolor{gray!20}
         \cellcolor{white}{} & \multicolumn{2}{l}{Publish probes IP address list for opt-out.} & \benign & \evil  & \evil  & \evil \\
       {Transparency\tnote{1}} & \multicolumn{2}{l}{Use fixed IP addresses instead of trashable ones.} & \benign & \benign & \evil & \evil\\
        \rowcolor{gray!20}
         \cellcolor{white}{} & \multicolumn{2}{l}{Set whois records with organization and abuse email.} & \benign & \neutral & \evil & \evil \\ 
         % \rowcolor{gray!20}
         & \multicolumn{2}{l}{Reverse DNS pointing to the company.} & \neutral & \neutral & \evil & \evil \\
         % \rowcolor{gray!20}

        \midrule
     %    \multirow{4}{*}{Harmlessness\tnote{2}} &  \multirow{4}{*}{Unauthorized access.} & {IoT (IP Camera)} & \benign & \evil & \benign & \benign \\
     % & &{IoT (OpenWrt Router)} & \benign  & \benign & \benign & \evil \\
     % &  &{Web (Prometheus)} & \evil & \benign &  \benign & \benign \\
     % & &{Web (Elasticsearch)} & \benign  & \benign & \evil & \benign \\
     \rowcolor{gray!20}
      \cellcolor{white}{} & {Malformed requests} & ~& \benign  &  \benign &  \benign & \evil\\
        % \midrule
        \cmidrule(l){2-7}
        % Privacy
        ~  &\textbf{Unauthorized Access Service}  & \textbf{Minimized Probe} \\ 
         \cmidrule(l){2-7}
        \rowcolor{gray!20}
         \cellcolor{white}{} & FTP &  Null Probe & \benign  & \evil  & \evil  & \evil \\ 
        & Redis  & Command: ping & \evil  & \evil  & \evil  & \evil  \\ 
        \rowcolor{gray!20}
       \cellcolor{white}{}  &ZooKeeper  & Command: ruok & \evil  & \evil  & \evil  & \evil  \\ 
        & ElasticSearch  & Path: / & \evil  & \evil  & \evil  & \benign  \\ 
        \rowcolor{gray!20}

        \cellcolor{white}{Harmlessness\tnote{2}} & MongoDB  & Command: mongo& \evil  & \evil  & \evil  & \evil  \\ 
         & RDP  & RDP Handshake & \benign  & \evil  & \evil  & \evil  \\ 
        \rowcolor{gray!20}
        \cellcolor{white}{} &LDAP  & LDAP Handshake & \evil & \evil & \benign  & \benign  \\ 
        & Memcached  & Command: stats & \benign  & \evil  & \benign  & \benign  \\ 
        \rowcolor{gray!20}
        \cellcolor{white}{}& CouchDB  & Path: / & \benign  & \evil  & \evil  & \benign  \\

        &{IP Camera(Web Service)}   & Path: / & \benign & \evil & \benign & \benign \\ 
        \rowcolor{gray!20}
        \cellcolor{white}{}&{OpenWrt Router(Web Service)} & Path: / & \benign  & \benign & \benign & \evil \\
        &{Prometheus(Web Service)}  & Path: /  & \evil & \benign &  \benign & \benign \\
        \midrule

        % Anonymization
        % \rowcolor{gray!20}\cellcolor{white}{Anonymization} \multirow{10}{*}{\cellcolor{white}{Anonymization}}
        \rowcolor{gray!20}
        \cellcolor{white} &  FTP  & ~ & \neutral  & \neutral  & \neutral  & \neutral \\ 
        &Redis&   & \neutral  & \evil  & \neutral  & \neutral \\ 
        \rowcolor{gray!20}
        \cellcolor{white}{}&ZooKeeper  & ~ & \neutral  &  \neutral  & \neutral  &  \neutral  \\ 
        &ElasticSearch & ~ & \neutral  & \evil  & \evil  & \neutral \\ 
        \rowcolor{gray!20}
        \cellcolor{white}{Anonymity\tnote{3}} & MongoDB  & ~& \neutral  & \evil  & \evil  & \neutral \\ 
        &RDP & ~ & \benign  & \evil  & \evil  & \evil  \\ 
        \rowcolor{gray!20}
        \cellcolor{white}{} & {LDAP} &  & \evil & \evil & \benign  & \benign  \\ 
        &Memcached &   & \neutral  & \neutral  & \neutral  & \neutral \\ 
        \rowcolor{gray!20}
        \cellcolor{white}{} & {CouchDB}  & ~ & \neutral  & \evil  & \evil  & \neutral \\
        &IP Camera  & ~ & \benign & \evil & \benign & \benign\\

        \bottomrule
    \end{tabular} 
    \begin{tablenotes}  
            % \footnotesize
            \item[1] {\noteBenign} indicates scanners obey guidance, {\noteNeutral} indicates scanners obey the guidance partially, and {\noteEvil} indicates all scanner violate transparency principle. 
            \item[2]  {\noteBenign} indicates that the engine only sends standard and minimized probes, and {\noteEvil} indicates the use of malformed or infiltrated requests.
            \item[3] {\noteBenign} indicates that PII has been fully anonymized, {\noteNeutral} indicates only the assist software version, which may facilitate attacker infiltration, and  {\noteEvil} indicates sensitive PII has not been anonymized and leaked.
    \end{tablenotes}  
    \end{threeparttable}
    % }
\end{table*}

% 知情权 去掉5和6
\subsection{Transparency}\label{sec:ethical_guideline}

% Regardless of the country, personal information privacy laws universally require that the collection of personal data must be conducted with the knowledge and consent of the individuals involved. These laws emphasize transparency, ensuring that individuals are fully informed about what data is being collected, how it will be used, and who will have access to it.
In search engine crawler standards, transparency about crawler identity is crucial, since crawlers are required to clearly inform users about data collection practices and purposes, meanwhile, users can protect opt-out rights by \textit{robots.txt}.
As \engines cover a broader scope than search engines, they also bear responsibility for clear disclosure to signal benign scanning intent. We summarized the five best actions for transparency. 
In total, Censys and Shodan have made conscious efforts to make their identities and activities transparent to users, while FOFA and ZoomEye are not.

% While there isn't an official standard for \engines, notable tools and engines such as ZMap, Censys, and Onyphe have introduced some best practices~\cite{durumeric2013zmap,durumeric2015search,onyphe-standard}. 

% Hence, we explore whether the \engines adhere to the guideline proposed by Onyphe\geng{why Onyphe, instead of Zmap?} in 2022.
% 写我们是结合他俩summarize出来的

% performance -> action
% We summarized the ethical scanning performance\geng{performance?} of each \engine in Table~\ref{tab:ethic_guideline}.

\noindent\textbf{Explain the purpose of every probe.}
Network administrators may be wary of unauthorized scans, but if they understand the purpose is to identify vulnerabilities and offer security recommendations, they are more likely to permit such activities.
The best practice involves hosting a website on port 80 of each \scanip to describe the purpose and nature of the scan, recommended by three tools/engines that proposed best practices. As an alternative approach, declaring identity in the User-Agent header during HTTP scans can also signal intent, however, only HTTP scanning can be informed.

Unfortunately, testing revealed that none of the \scanips from the four engines provided such information, even Censys said it in its paper~\cite{durumeric2015search}.
% 
% ZMap\geng{?} recommends hosting a website on port 80 of each \scanip to describe the purpose and nature of the scan. However, testing revealed that none of the \ scans from the four engines provided such information, even if ZMap and Censys are built by the same team.
% 
Specifically, we found that 83 FOFA IPs and 9 ZoomEye IPs had port 80 open, but the content varied significantly, ranging from nginx test pages and device login pages to WordPress websites, indicating that these IPs may be resold and used by others.

% 尽管censys提出了一些xxx，但是没有任何一个厂商在implementation里干了，包括censys自己
\finding{Although Censys proposed the best practice of hosting a website on port 80 of each ScanIP to describe the scan's purpose and nature, no engines, including Censys itself,  fully follow it in implementation.}

% Another approach is to declare one's identity in the User-Agent header when scanning HTTP services. This method helps \cite{li2021good} differentiate search-engine bots, and academic and industry scanners from malicious bots.
% 
Only Censys identifies itself as ``Mozilla/5.0 (compatible; CensysInspect/1.1; +https://about.Censys.io/)'' in User-Agent, with 31\% of scans targeting the root directory lack a UA. Other engines claim to be users of Chrome or Firefox on Windows, Linux, or macOS, as shown in Table~\ref{tab:useragent}.

\noindent\textbf{Publish scanner IP address list for opt-out.}
To respect users' privacy and information security, an opt-out option should be provided, as required by major privacy regulations~\cite{gdpr,CCPA,JapanPersonalInformationProtectionLaw}.
Unfortunately, only Censys offers explicit instructions for opting out of scanning activities.
We observed that FOFA responds aggressively to users who do not want to be scanned, advising them not to place their devices on the external network~\cite{fofa_help}. 
% This reflects a lack of user-orientation.

Censys takes a proactive approach by publishing IP ranges and suggesting users block their access via firewalls. They also inform users about filtering scans using the User-Agent. This transparency reflects Censys' commitment to ethical scanning practices and respect for user privacy.

\noindent\textbf{Use fixed IP addresses instead of tractable ones.}
As we discovered in Section~\ref{sec:ip_usage}, both FOFA and ZoomEye rotated their \scanips in one year, with FOFA specifically replacing its IP pool every three months. This practice poses a challenge for users attempting to evade scans by configuring their firewalls.

\noindent\textbf{Set whois records with organization and abuse email.} 
This helps users to easily identify and contact the engines in case of any abuse or issues related to their IP addresses.
Among the four engines, Censys is the only one that set its \scanips with its own abuse email and organization. 
In contrast, FOFA and ZoomEye both utilize the whois information of their cloud service providers, rather than maintaining their own information. 
Shodan, similarly, only has an IP segment with 11 \scanips with whois information associated directly with Shodan, while 81.7\% of its ScanIPs use the Whois details provided by its cloud service providers.

\noindent\textbf{Reverse DNS pointing to the company.} 
Among the four engines, only Shodan and Censys have reverse DNS records associated with their scanning IPs. Shodan's reverse DNS records point to \url{scanf.shodan.io} or \url{census.shodan.io}, while Censys' records point to \url{censys-scanner.com}. Notably, there are 23 \scanips from Shodan and 24 \scanips from Censys without corresponding reverse DNS records. Interestingly, the IPs lacking RDNS records from Censys are within their publicly announced IP ranges. Regarding Shodan, the \scanips come from the same subnet, suggesting their association with Shodan. In contrast, neither ZoomEye nor FOFA assigns reverse DNS records to their \scanips.

% \finding{Users cannot discern, with no possibility whatsoever, whether the scanning originates from FOFA or ZoomEye through IP homepages, WHOIS, Reverse DNS, or public listings.}
\finding{Through the analysis of 1,407 \scanips, users cannot identify whether the scans originate from FOFA or ZoomEye. This makes users hard to identify and evade scanning.}
% \engine ecosystem highly opaque.}
% including details such as IP homepages, WHOIS, reverse DNS, or public listings
% opaque 不透明的 没有找到transparent加个in/un/im/non之类的词

% \noindent\textbf{Handle abuse requests and remove collected data requests on a timely manner, ask no question.}

% intrusive 把第一段的5和6搬过来
\subsection{Harmlessness} \label{sec:sensitive_path} 
Cybersecurity-related laws protect computers attacks, including intentionally accessing a computer without authorization or exceeding authorized access to obtain information or recklessly causing damage. \Engines ensure that their scanning activities are harmless, such as only sending standard requests and accessing permitted resources.
However, we observed harmful scanning in our honeypots, from all \engines, including malformed requests and attempts at unauthorized access, which may lead to system error, data breaches, or vulnerability exposure.

% \noindent \textbf{Scanning paths.} Table~\ref{tab:web_path} shows the default paths scanned by each \engines during web scans. Both the root path (/) and the icon (/favicon.ico) are commonly requested by all engines, as they often serve as indicators of the web server's existence and functionality.
% % 
% 

% \noindent\textbf{Inaccessible paths.}
% The website administrator can define accessible and inaccessible paths to protect certain areas from unauthorized access. 
We first investigate the default scanning paths of them, as shown in Table~\ref{tab:web_path}. All engines commonly request the root path (/) and the icon (/favicon.ico), which indicate the web server's existence and functionality.
Additionally, Shodan and ZoomEye access robots.txt, security.txt, and sitemap.xml files, providing supplementary website information. Notably, security.txt files provide pathways for reporting security issues, facilitating communication between researchers and administrators, and highlighting the engines' proactive role in enhancing cybersecurity practices.
% In our study, we customized paths commonly scanned by search engines, like robots.txt and sitemap.xml, and marked 12 vulnerable paths as inaccessible. 
% Gladly, although the engines accessed these files, they did not attempt to access the specified paths within them.

\subsubsection{Malformed requests}
Sending malformed data packets or protocol requests can potentially lead to abnormal behaviors in target systems or network devices. We found that ZoomEye, by default, employs a malformed probe in the form of ``GET /nice\%20ports\%2C/Tri\%6Eity.txt\%2ebak HTTP/1.0\textbackslash r\textbackslash n\textbackslash r\textbackslash n'' for all web services, which we can decode to a more friendly ``/nice ports,/Trinity.txt.bak''.
This probe comes from Nmap's service detection~\cite{nice-port}, uses ASCII escaped characters in an attempt to generate an HTTP 404 error message to probe a web server, which is one of the top four web services exploits in 2019~\cite{top10exploits}.
% The design intent of this request lies in leveraging potential errors or exceptional responses generated by web servers when processing specific requests to gather information. 
A successful scan can reveal crucial details about the web server's codebase and potentially even expose vulnerabilities through response headers and error messages. As a result, this technique is often exploited by attackers. 

\subsubsection{Unauthorized access} 
Unauthorized access involves bypassing security measures, exploiting vulnerabilities, and leveraging weak authentication.
An ethical \engine should adhere to data minimization principles during scanning, avoiding unauthorized access to sensitive paths on a user's host to prevent potential privacy breaches.
% as this could inadvertently facilitate privacy breaches by potential attackers.
% However, based on the data collected by our web honeypot, we found that all four \engines scanned unauthorized paths, aiming IoT devices and databases, as detailed in Table~\ref{tab:censys_sensitive_path}.
Specifically, these engines attempt to access paths requiring authentication but are often left insecure. This behavior indicates that some engines may view user data as a key component of their commercial value, without users' knowledge or consent.

\noindent\textbf{Minimized scanning. }
% 什么是infiltration，定义default和infiltration的区别
To clarify whether engines acquire data unethically when finding a service, we first define the minimized actions and the infiltration actions in scanning.

% 谨防有人看不懂(我的中文就很绕了，不确定我改出来的英文是不是很绕，反正gpt被绕晕了）：given that测绘引擎将提供服务标签作为他们功能的一种，我们遵循data minimisation原则，将识别服务的标准操作定义为只要能够确认该端口上是该服务就应停止。
Given that \engines provide service tags as part of their functionality, we define the minimized action as probing to confirm a service on a port and ceasing further scanning, aligning with data minimization principles.
In contrast, infiltration probes aim to get more detailed and private service information once a service has been identified.

For instance, a minimized probe, i.e., \textit{``GET /''} request is enough to verify ElasticSearch web server. However, using ``/\_cat/indices'' will over-collect database indices. 
Besides, some services, such as MongoDB, connecting successfully with specific tools can confirm the existence of the MongoDB service. 
Any subsequent interaction probes are considered infiltration.
Similarly, fetching the FTP welcome banner after the handshake can know an FTP server, just like opening a webpage. Attempting anonymous FTP login is like infiltrating the webpage's login system, exposing weakly protected hosts and aiding attackers in identifying potential victims.

We selected ten common services vulnerable to unauthorized access and deployed them as response templates.
We used interaction tools to probe and understand the requisite actions needed to elicit various responses and then defined the minimized probe for each service.
We found that some services even do not require specific clients for confirmation, for example, the \texttt{PING} command is sufficient to determine the presence of Redis on a host, evidenced by the response ``PONG''. Also, send \texttt{ruok} to ZooKeeper will receive \texttt{imok}.

We use engine records and traffic captured by our honeypots to determine whether the engines attempt unauthorized access to infiltrate services.
% To gain further insight into whether the acquisition of private information by the four \engines is reasonable, 
% we analyzed their record for some services which are vulnerable to unauthorized access. 
%
Specifically, we searched the host records containing these services in each \engine and manually inspected the first 100 entries to check for potential excessive data acquisition, based on our defined minimized probe. 
For web services, we learn from our honeypots.

\noindent\textbf{Infiltration of \engines. }
The result in Table~\ref{tab:ethical violation} shows that unauthorized access is widely attempted among the \engines. Six services are infiltrated by at least three engines. 
% including Redis~\cite{redis_protocol}, ZooKeeper~\cite{ZooKeeper}, MongoDB~\cite{MongoDB_protocol}, FTP~\cite{ftp_rfc}, ElasticSearch~\cite{Elasticsearch} and RDP~\cite{rdp_protocol}. 
Engines connect and then send additional commands.
% anonymously log in, and send querying commands to get excessive system details, list the data entries, and enumerate the databases.
This includes anonymously logging in to FTP, getting system details of ZooKeeper(\texttt{stat}) and RDP, and enumerating the databases of ElasticSearch(\texttt{/\_cat/indices}), MongoDB(\texttt{show dbs}), and Redis(\texttt{keys *}).

The probe, leveraging by three engines (except for Censys), used for RDP is exploiting a vulnerability with a CVSS3 score of 9.8~\cite{rdp-ntlm-info}. The script \texttt{rdp-ntlm-info} in Nmap sends an incomplete CredSSP (NTLM) authentication request with null credentials, which causes the remote service to respond with an NTLMSSP message of CredSSP (NTLM).

Successful infiltrations exposed weakly protected hosts lacking authentication. For instance, 
% while the \texttt{PING} command is sufficient to determine the presence of Redis on a host, evidenced by the response ``PONG'', the utilization of the INFO command has uncovered a significant security concern. This command, when executed on unauthenticated hosts, grants unauthorized access to Redis information, 
the non-error response of the INFO command on Redis hosts granted unauthorized access to Redis information, revealing that 74.97\%(59,725/79,664) of Redis hosts listed on Shodan and 182,137 hosts on Fofa are vulnerable to arbitrary access. Similarly, the success in logging into FTP service exposes 135,599 FTP hosts listed in ZoomEye do not need authentication. Only Censys, who does not infiltrate FTP host, will not tell the authentication information.
% 
% 可以陪一段ZooKeeper，可以只是用客户端链接，但打了个stat命令后99%的未授权都被暴露了
The issue surrounding ZooKeeper is even more critical. While a \texttt{ruok} request or client handshake can confirm the service's existence, the status response to the ``stat'' command reveals 99.91\% (369,552/369,881) of the recorded hosts are vulnerable to unauthenticated access, with only 0.09\% of the hosts resisting these probes.

% 顺序：最严重的-》最意外的-〉另外两个-》有意思的case
Shodan is the most serious among the four engines, almost infiltrating all services.
% Shodan reveals a number of PII including client IP addresses stored in ZooKeeper, and telephone numbers and emails found in LDAP.
% At the same time, Shodan uses additional probes to obtain more detailed service information based on domain knowledge, such as MongoDB\cite{MongoDB_protocol}, Memcached\cite{Memcached}, and Remote Desktop Protocol(RDP)\cite{rdp_protocol}. What's even worse is, 
% Shodan tries \ww{logging into FTP as an anonymous user, and enumerating} the contents of databases such as CouchDB~\cite{CouchDB_protocol} and Redis~\cite{redis_protocol}. 
Memcached~\cite{Memcached}, and IP Cameras are only infiltrated by Shodan. 
% After connecting to a Redis database, Shodan proceeded to inspect keys and connected client information within the database. 
For Memcached, Shodan sent a ``stats'' command followed by ``stats settings'' to retrieve additional configuration information.
Our honeypots detected that Shodan attempted to access and retain 25 sensitive paths for IP camera configuration details and real-time feeds, violating user privacy as outlined in Table~\ref{tab:sensitive_path}. Notably, some hosts provided by Shodan did not offer real-time images in their root directories, suggesting Shodan probed deeper paths, confirming our honeypot findings.
Also, this helps the attackers locate and exploit IP cameras that are accessible without authentication, who can abuse it for illicit camera spying and exacerbating the sale of voyeuristic content. 
Evidence of this trend lies in Shodan's Explore module, where seven of the top 10 voted queries focus on seeking live webcam feeds, with one even titled ``live sex cam''.

Although Censys claims that they never try to log into any service, read any database, or gain authenticated access to any system, we still find Censys infiltrated six services, 
such as getting server detail of ElasticSearch(\texttt{/\_nodes}) and MongoDB(\texttt{isMaster} and \texttt{Buildinfo}).
Also, Censys extracts the user's information including email, company, department and telephone in the Lightweight Directory Access Protocol (LDAP)~\cite{ldap_rfc} service.
As for web services, Censys scanned nine paths of the Prometheus server, which facilitates monitoring system metrics and alerting. However, Censys stated that it introduced granular recognition for Prometheus in 2019~\cite{censys_prometheus}, allowing users to search for exposed Prometheus endpoints.

Besides,
% FOFA and ZoomEye also attempt to use some infiltrate probes to get additional service information including RDP and MongoDB, and to log into FTP services as anonymous users. 
Zoomeye enumerates files using \texttt{LIST} command after logging in FTP, and accesses the sensitive path ``/cgi-bin/luci/'' of OpenWrt routers,
% LuCI (Luci embedded web interface) is a web-based user interface for the OpenWrt router system, 
a web interface that allows users to configure and manage the router through the browser. 
FOFA exhibits a preference for enumerating indices and extracting database information within database-like services, such as ElasticSearch~\cite{Elasticsearch} and CouchDB~\cite{CouchDB_protocol}. We also witness FOFA acquiring ``/\_cat/indices'' for ElasticSearch in our honeypots.

% By accessing the ``/cgi-bin/luci/'' path, attackers can open the LuCI interface and carry out operations such as network settings, firewall configuration, and software management.
% 如xx表所示，我们发现Censys会扫描Prometheus服务器的路径。Prometheus作为一种流行的系统监控工具，允许用户收集、存储并对其环境中系统的指标进行警报。然而，Prometheus默认情况下没有身份验证。没有适当的访问控制，任何拥有互联网访问权限的人都可以查看Prometheus实例中的活动。有关组织系统和设备的信息对于执行针对组织的侦察的威胁行为者可能非常有用。
% 尽管如此，我们发现Censys在2019年就已经将该协议添加到它们的IPv4数据集中，从而用于帮助用户搜索发现暴露的Prometheus端点。

% Consequently, threat actors may potentially glean valuable information about an organization's devices and systems, intensifying the preliminary phase of a cyberattack against the organization. 
% These paths are all unauthorized paths, however, only Censys stated that they introduced granular recognition for Prometheus in 2019~\cite{censys_prometheus}, allowing users to search for exposed Prometheus endpoints. 
% This capability is instrumental in mitigating security risks associated with inadvertently accessible but inactive endpoints on the Internet.
% Despite FOFA claiming to us that they never scan with paths, they did not provide a direct response when we inquired about the purpose of accessing this path. However, we confirmed in Section~\ref{sec:privacy_data} that FOFA indeed accessed this path to display index information for each Elasticsearch cluster in their search results.
% 
Interestingly, in our communications with these \engines, they all claimed to have only scanned the root directory or paths like robots.txt, explicitly denying any scanning of sensitive paths. However, when asked about discrepancies between their behavior and claims, they refused to answer and ended communication.
% However, since we observed 10-58 IPs in our honeypots scanning every sensitive path mentioned above, we can confirm that these behaviors are not false positives introduced when we were looking for \scanips. 
% Also, we confirmed they indeed accessed this path as they display this information in their search results, as experimented in Section~\ref{sec:privacy_data}.

% In conclusion, in order to obtain more detailed service information, all \engines will inevitably use some infiltrate probes. Specifically, Shodan, FOFA, and ZoomEye even try to log into unauthorized systems. Moreover, to showcase its comprehensive capabilities, Shodan even provides screenshots of the services separately.

% infringement 侵权

% \noindent\textbf{Non-authentication disclosure.}
% To evaluate the severity of the harm caused by disrespectful data access, we analyzed the number of vulnerable hosts impacted by ten services across various engines. 
% 
% Our findings reveal that \ww{the infiltration scanning} employed by engines has led to the 

% In addition to exposing that the hosts allow arbitrary access, we found that 48 CVEs of the 10 services require unauthenticated access as an entry, leading to arbitrary code execution and denial-of-service attacks.

\ignore{
\finding{
% FOFA, ZoomEye, and Shodan were found to engage in scanning unauthorized access, contradicting their claims of harmless scanning strategies.
The successful infiltration of the engines additionally exposes the hosts that lack authentication, a vulnerability that could have been avoided.
}}

\finding{
% The asset investigation conducted by \engines to provide extensive service details based on widespread disrespectful scanning.
% The \engines widely infiltrate services without proper authentication to get extensive details.
The \engines send malformed requests, attempt to access excessive details without authorization, and even exploit vulnerabilities, posing risks to user privacy and security.
}

\subsection{Anonymity}\label{sec:anonymity}

Anonymity refers to hiding personal information when displaying search results. The privacy laws require that published data cannot identify specific natural persons and cannot be reversed or reconstructed, to prevent user data from leakage.
% result 展示的对不对
% 由于隐私数据种类多样，定义也各有不同，本文中将服务信息中的个人可识别信息、数据库信息和软件版本信息定义为隐私数据。
Even when engines use minimized probes in Section~\ref{sec:sensitive_path}, certain responses can still contain private information. Failure to anonymize the privacy before displaying on search results can lead to privacy leakage risks.
% and to some extent help the privacy trafficking industry.

\noindent\textbf{Privacy data in device search engine result.} 
Due to the variety of types of private data and their different definitions, we specifically focus on privacy data in the ten services' responses,
 % as privacy data\ww{personally} identifiable information (PII), database information, and software version information
% 
includes host names, user names, avatars, emails, geographical location, screenshots, \etc Such information might be abused by network attackers, such as launching social engineering. In addition, the leakage of database information may lead to theft of valuable user information.

Leaking software versions has caused huge risks. The OWASP top 10: 2021~\cite{owasp2021} highlights ``Vulnerable and outdated components'', indicating that many hosts have not been updated to the latest version and remain susceptible to security threats. 
% As for the software version, 
% it is widely used by security administrators to monitor and assess the impact of vulnerabilities. 
% 
What's more, security companies~\cite{turingsecure,invicti,smartscanner} treat version disclosure as a vulnerability, as attackers can exploit known vulnerabilities associated with disclosed versions.
% Also, identifying the version is a key premise for a supply chain attack.
% Although updating to the newest version can prevent attack,
% 
% However, the version can tell whether a host is able to be attacked or not, we found that xx CVEs of the 10 services associated
% with specific software versions. If the software version number is disclosed on the engine, attackers can easily identify and exploit these vulnerabilities to target the host.
% 
% Therefore, software versions should be sensitive information.
% and treated with the appropriate level of privacy protection.
% 

\noindent\textbf{Privacy leakage.}
Here we use the same assessment method with Section~\ref{sec:sensitive_path}, and manually inspect the first 100 search records of 10 services.
Considering that engines may not categorize software versions as privacy and widely publish this information, we isolated it in our experiments to avoid influencing the results of other privacy leaks, using a yellow face in Table~\ref{tab:ethical violation}.
% 
% The result shows that all infiltration probes lead to privacy leakage. \geng{1 or 2 sentences to summarize the result first.}
% Shodan and FOFA do not prioritize user privacy. Specifically,
The result shows that  Shodan exposed database or PII for 7 out of 10 services. Furthermore, version information is widely exposed in the records.

Database services' data indices are being exposed and displayed, as intentionally customized features tailored for these services. 
We found 145,310 database indices of Elasticsearch, 178,879 indices of MongoDB, and 2,306 databases of CouchDB are showing on Shodan and FOFA. What's more, Shodan lists 68,543 Redis hosts with their keys.

What's worse, Shodan provides an image display platform on \url{https://images.shodan.io/}, displaying images of IP cameras they discovered and log-in screenshots of RDP services with the avatars and usernames. 65,042 camera snapshots and 791,333 remote desktop screenshots are displayed upon submission of this paper.
% \ww{We have inquired Shodan about the social significance of showing images and have not received a response yet.}

In contrast, Censys strives to mitigate privacy risks by displaying only relevant fields from responses, successfully avoiding leaking any private information on RDP and ZooKeeper. However, they still inadvertently leaked 230 LDAP user data (including name, email, company, address, and phone), along with Elasticsearch node and MongoDB device configurations.

ZooKeeper~\cite{Zookeeper} typically reveals all connected clients by default. Notably, FOFA masks all client IP addresses when displaying ZooKeeper results, while ZoomEye only masks its own IP. This highlights the different approaches various engines take in handling sensitive information.

\finding{ The \engines fail to anonymize asset sensitive data, including PII~(735 phone numbers, 65,042 cameras, 791,333 remote desktop screenshots, \etc), 326,495 database index and entries, before publishing on their search results.}

\ignore{
\wmy{an experiment assess the cves related to version}
\ww{\noindent\textbf{Version leakage risk.}
The risks associated with PII leakage are well-documented, but the dangers of version leakage are less widely recognized. We demonstrate the risks of version leakage by examining the number of vulnerabilities associated with specific software versions. If the software version number is disclosed on the engine, attackers can easily identify and exploit these vulnerabilities to target the host.
}}

\section{Discussion}
\subsection{Suggestions}
By uncovering the scanner IPs of \engines, our original findings expose significant ethical considerations in the engines' scanning activities, \ie lack of transparency, harmlessness, and anonymity. Our findings underscore the pressing need for stringent ethical standards and regulatory oversight in the use of these engines. 
% Therefore, based on previous guidelines, the Menlo Report, and user perspectives, we propose the following recommendations.
Therefore, based on the behavior of \engines, we propose the following suggestions for both users and engines.

To avoid being scanned, users can use WHOIS and reverse DNS records to find and block IPs from transparent engines. For those engines that do not use fixed IPs, users can leverage public blocklists such as AbuseIPDB, as 47.26\%(665/1,407) scanner IPs we found have been reported and labeled. Users can also report suspicious scanning IPs to help others. 
If users have to expose services on the public network, we recommend concealing them by migrating default ports to random ports, rather than neighbor ports or the ports we show in Table~\ref{tab:service_multiport}.

Our research reveals a substantial number of unauthenticated services exposed to attackers due to excessive infiltration by the \engines. Users should check if their services are left unauthenticated, as we have found that at least 48 CVEs associated with 10 services require unauthenticated access as an entry point, leading to potential risks such as arbitrary code execution and denial-of-service attacks.
% 

% From a security perspective, scanning for vulnerable services should be conducted in a manner that ensures sensitive information, including software versions, is reported to the 

We suggest that device search engines enhance their ethical scanning practices. To improve transparency, they should clearly explain the purpose of each scan and provide users with an opt-out option by publishing a list of scanner IP addresses. Additionally, using fixed IP addresses instead of disposable ones can further improve trust.
Moreover, device search engines should minimize potential harm to hosts by sending only standard, minimal probes and should refrain from exploiting any vulnerabilities or unauthorized services. They must also avoid excessive probing of user devices to enhance functionality, particularly when it involves accessing private data.
Finally, to protect user privacy when displaying data, engines should anonymize any potentially sensitive information. We recommend that engines report vulnerabilities and privacy leaks to the appropriate stakeholders rather than disclosing them publicly in search results or other open channels.
% To enhance transparency in scanning practices, it is recommended to clearly explain the purpose of each probe conducted and provide users with an opt-out option by publishing a list of probe IP addresses. Additionally, using fixed IP addresses instead of disposable ones maintaining accurate WHOIS records with organizations, and abusing contact email information can contribute to improved transparency. Furthermore, ensuring that reverse DNS records accurately point to the company's domain can enhance accountability and transparency in scanning activities.

% \ww{\Engines} should avoid potential harm to the hosts, by sending standard and slow requests. Also, any vulnerabilities or unauthorized services should not be exploited. 
% 
% Similarly, \engines should refrain from excessively probing user devices to enrich their functionality, especially when it involves accessing users' private data. 
% 
% \ww{\Engines} should also ensure the protection of user privacy when displaying data by anonymizing possible sensitive information, as the scope of device discovery should not include the discovery of private data on devices.

\subsection{Limitations}
Although our work has uncovered 1,407 \scanips and their strategies, we have some limitations.
First, since our \scanips are collected based on TCP services, any \engine that differentiates between TCP and UDP scanning may miss \scanips dedicated solely to UDP services and their corresponding behaviors.
Also, our method only captures exposed \scanips; many scanner IPs may remain undiscovered if they have not scanned easily exposed services.

Secondly, the artificial nature of honeypots may bring potential bias. Honeypots are commonly used in cybersecurity research, such as \cite{sasaki_exposed_2022}, \cite{srinivasa2021open}, \cite{cetin_cleaning_2019}), reflecting real-world attack patterns and offering valuable insights into network threats. 
To minimize potential bias, our honeypots are deployed and configured to closely mimic real system behavior, reducing discrepancies with real-world environments. For example, we decoy camera snapshots with dynamic timestamps to simulate real-time monitoring scenarios. Additionally, we deployed honeypots across multiple countries and collected data over a year to ensure diversity and representativeness.

However, given our honeypot number and monitoring period,
% Secondly, due to our limited number of honeypots and the one-year monitoring period, 
certain aspects may lack statistical conclusions, such as scanning preferences across geographical regions and the periodicity of full-port scans. We believe that deploying more honeypots over an extended duration may give the conclusions.

Due to limitations in computing resources and manpower, our understanding of \engines' behavior is primarily focused on web services representative of IoT devices. Expanding the services simulated on our honeypots could provide insights into a broader range of engines' behaviors. For services beyond the web, we also manually examined their privacy leakage behaviors in Section~\ref{sec:anonymity}.

% \subsection{Compare to search engine}
% As the \engine is similar to search engines, one may wonder how the ethical performances of the search engine bots. Due to we use new IPs and domains for our honeypots and never advertise to human users, we can hardly receive traffic from search engines. However, previous works~\cite{sun2010ethicality} have analyzed the ethics of search engines by investigating how web crawlers respect the regulations set forth in robots.txt configuration files. They proved that most commercial web crawlers' behaviors are ethical, but still consistently violate or misinterpret certain robots.txt rules.

\subsection{Ethics and Disclosure}
\noindent\textbf{Ethic concerns.}
In our research, we adhere to ethical guidelines by utilizing publicly available data provided by device search engines to locate their \scanips, without engaging in any database attacks against these engines. Additionally, during our periodic searches for IP Mirror Services, we strictly abide by the limitations of our purchased membership account, responsibly using their query API.

\noindent\textbf{Disclosure.}
During the process of collecting traffic in our honeypots, we actively engage with device search engines.
%to ensure that our interesting findings are not a result of individual incorrect \scanips. 
Each engine responded positively when we reported suspicious scanning IPs.
For example, upon discovering 14 IPs using Censys' user agent but not listed in Censys' public IP range, we promptly reported this to Censys and confirmed that they were indeed fake. Additionally, we encountered instances where attackers exploited discarded \scanips from FOFA and ZoomEye to conduct malicious scans against us. After reporting these incidents, both FOFA and ZoomEye confirmed that the IPs in question had been abandoned.

\section{Conclusion}

This study presents the first comprehensive assessment of the assets, operational strategies, and ethical concerns of device search engines, providing original findings rather than reiterating existing information. Through innovative methodologies, we collected 1,407 scanner IPs and demonstrated that users could hardly evade scans by blocklisting scanner IPs or migrating service ports. Our research exposes significant ethical breaches—primarily the lack of transparency, harmlessness, and anonymity in their scanning activities. 
% \ww{This study presents the first comprehensive examination of these engines' operational and ethical dimensions and gains deep original findings instead of reiterations of information already stated by the \engines.} 
These findings underscore the pressing need for stringent ethical standards and regulatory oversight in the use of these engines, which are pivotal in network security but also pose risks to user privacy.

Given these issues, we advocate for the formulation of clear ethical guidelines and the establishment of robust regulatory frameworks to govern the operations of device search engines. 
In conclusion, while device search engines are invaluable for network security, their responsible use is paramount. Our study calls for a balanced approach that aligns the powerful capabilities of these tools with stringent ethical practices, thus protecting users and strengthening the security landscape.

% These should aim to enhance transparency, enforce respect for user privacy, and ensure that scanning activities are conducted in a harmless and respectful manner.
% 

\section*{Acknowledgment}
We would like to thank Bingjing Song for providing technical and data assistance for this paper. 
We also thank the anonymous reviewers for their insightful comments that helped improve the quality of the paper. This work was supported in part by National Natural Science Foundation of China (62302101). Min Yang is the corresponding author, and a faculty of Shanghai Institute of Intelligent Electronics \& Systems, and Engineering Research Center of Cyber Security Auditing and Monitoring, Ministry of Education, China.

% \normalem
\bibliographystyle{IEEEtranN}
% \bibliographystyle{ACM-Reference-Format}
% \bibliographystyle{IEEEtranS}
% \balance
\bibliography{reference}
\newpage
\appendix
\section{Methodology}
% 这个表表有必要吗？可以放appendix
\begin{table}[h]
    \centering
    \caption{Search syntax for search host 1.1.1.1:443 in different \engines.}
    \label{tab:engine_syntax}
    \begin{tabular}{c c}
        \toprule
        \textbf{Engine} & \textbf{Syntax} \\ 
        \midrule
        Censys & (ip="1.1.1.1") and services.port=`443` \\ 
        Shodan & ip:1.1.1.1 port:443 \\ 
        FOFA & ip="1.1.1.1" \&\& port="443" \\ 
        ZoomEye & ip:"1.1.1.1"+port:443 \\ 
        \bottomrule
    \end{tabular}
\end{table}
\begin{figure}[h]
    \centering
    \includegraphics[width=0.7\linewidth]{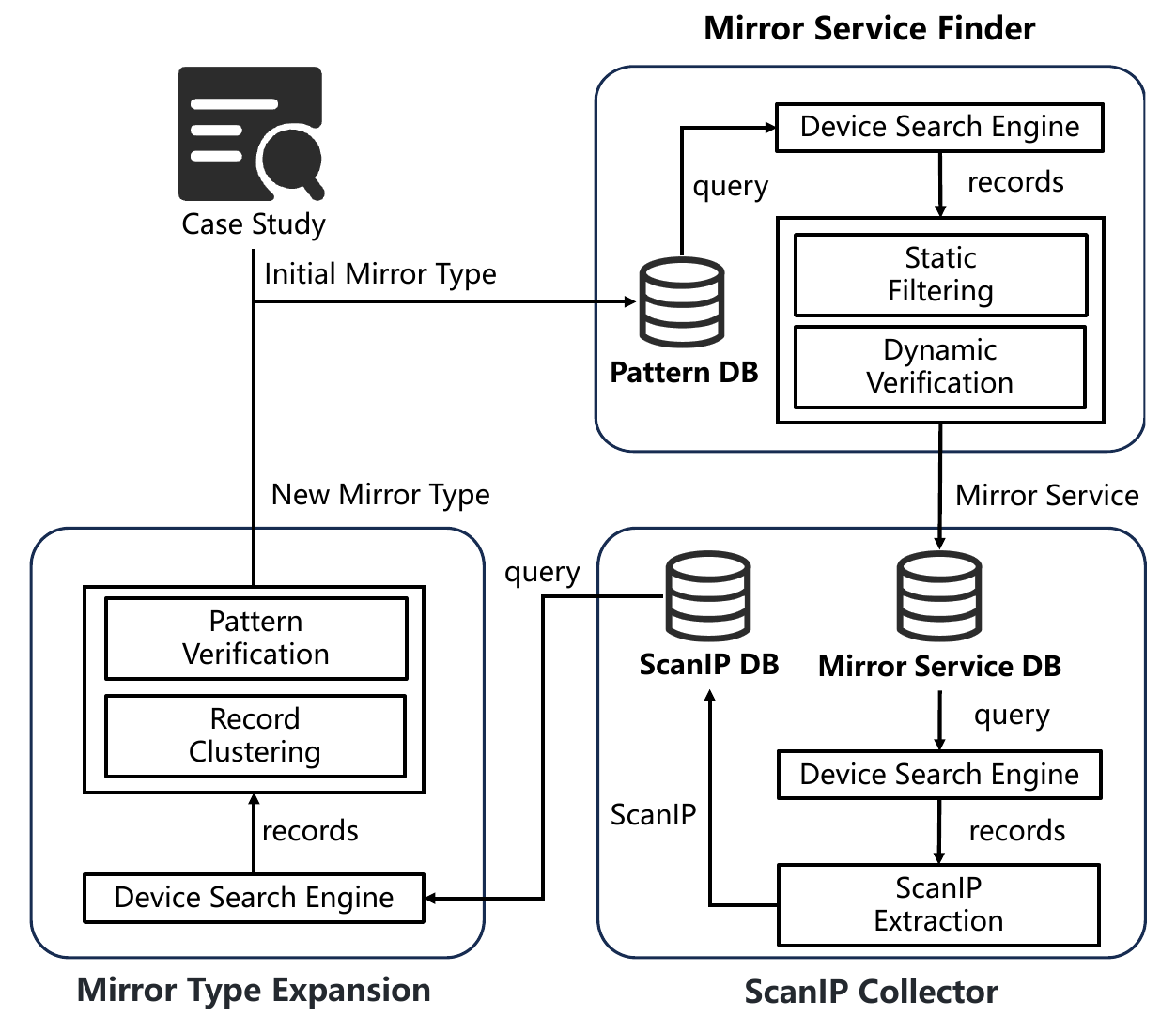}
    \caption{Overview of \scanip collection module.}
    \label{fig:ip_collection}
\end{figure}

\subsection{\scanip and \ipmirror Collection} \label{sec:implementation}
Based on our preliminary study, we selected four engines that offer sufficient queries and batch-automatable search API capabilities: Censys, Shodan, FOFA, and ZoomEye.
% 
% 因为互联网中的IP经常易主，搭载的服务也可能随时下线，由于需要与记录主机动态交互才可确定观测站，过早的数据记录中的虽然可能是测绘IP，但我们无法验证其真实性，因此我们限定使用每个平台中2023年3月-2024年3月之间的数据记录。
% 
To mitigate the impact of \scanip ownership changes and potential \ipmirror downtime, we limited our analysis to data records in \engines collected between March 2023 and March 2024.
% This decision was driven by the necessity of dynamic interaction with the host to determine whether it was an \ipmirror. While earlier data records may contain \scanips, their authenticity could not be verified.

% IPMirror Finder
% 实验过程中，我们发现有3家测绘引擎会尝试在记录中隐藏测绘IP。其中，ZoomEye和FOFA分别会将测绘IP替换成“xxx.xxx.xxx.xxx”、“*.*.*.*”的形式，而Shodan则会将扫描IP替换成224开头的D类多播地址。

% 基于上面不同测绘引擎的语法与功能特性，我们将pattern中的p_r转化成对应平台上的查询语法，获取符合要求的主机记录。接下来，我们使用python的re库从记录中匹配提取满足p_ip格式的candidate scanIP，并过滤掉IP地址不合法或者scanIP与服务器的IP地址一致的记录。然后我们使用不同协议对应的交互工具如nmap, nc, telnet, sipsak(sip交互工具)等，获取响应信息，并从响应中提取IP并判断真实性。
\noindent\textbf{Candidate \ipmirror Collection. }
% Firstly, we combined the service employed by \ipmirror with contextual keywords within its record to establish a pattern denoted as \textit{MirrorPattern} for identifying a specific type of \ipmirror.
We initiated our approach by extracting search patterns for \ipmirror identified in the preliminary study (Section~\ref{sec:preliminary}).
% 与ip和环境变量等可变因素无关的文本
We used the text in the records which are unrelative with the variable factors (IP and environment variables) and the protocol type as its pattern,
% leveraging protocol type and contextual keywords that don't include \scanip \geng{?} within its records, 
denoted as \textit{MirrorPattern}.
%拆成两句完整的话
% ippattern，mirrorpattern
For instance, the pattern for \ipmirror reflecting \scanip via the SIP protocol is defined as ``protocol:sip \&\& banner:received=''. Meanwhile, we designated the patterns that match the candidate \scanip in the responses as \textit{IPPattern}. For example, the \textit{IPPattern} corresponding to the SIP protocol is ``received=\$\{ipv4\}''. Then we defined the tuple \textless{}\textit{MirrorPattern}, \textit{IPPattern}\textgreater{} as a pattern capable of detecting \ipmirrors and mining \scanips.
We then converted \textit{MirrorPattern} into the corresponding syntax of the \engine and obtained matching host records by search APIs. 

% Leveraging the diverse syntax and functional attributes of the \engines, we transformed the MirrorPattern component of the pattern into the corresponding query syntax supported by each \engine and retrieved corresponding records. Subsequently, we employed the 're' library in Python to identify and extract candidate \scanips that corresponded to the IPPattern format from the records. Furthermore, we filtered out records containing invalid IPs or instances where the candidate \scanip coincided with the server's IP. 
\noindent\textbf{\ipmirror Verification.} After filtering invalid IPs and those that were the same as the host's IP, 
we used tools such as Nmap, Netcat, telnet, and sipsak\cite{sippak}, to probe the host and check if its response including our testing IP. If it does, we confirm it as a valid \ipmirror.

\noindent\textbf{\ipmirror Type Expansion.}
We took the \scanips we have collected as seeds and queried the records of the three types of scanned IPs on \engine. For instance, on ZoomEye, the query statement retrieving records whose responses contain the three variations of 1.2.3.4 is \texttt{banner: ``1.2.3.4'' banner: ``1\%2E2\%2E3\%2E4'' banner: ``4.3.2.1''}. 

% 
%我们获取的记录可能来自多种不同的服务，包括之前已经别发现的能够泄露IP的服务。因此，我们首先过滤满足已知服务pattern的记录。为了快速从剩下的记录中提取新的pattern，我们采用了聚类的方法。具体而言，由于测绘引擎记录中会提供协议信息，我们首先将不同协议的记录区分开。此外，由于记录中可能包含一些数字形式的可变信息（时间戳、服务版本号），这些信息可能会影响聚类效果，因为我们去除了响应记录中的数字信息。然后计算剩下的文件之间的相似度，将同一协议中文本相似度>0.9的作为同一类服务响应。
% 
Then, considering the records may come from various services, including those known \ipmirror types, we filtered out matching patterns of known \ipmirror types. For the remaining records, we utilized a clustering approach to extract new patterns for \ipmirror. Specifically, given that the \engine records provide service tags, we initially categorized the records according to their services. Then we recognized that these records could contain variable information in digital formats, such as timestamps or service versions. This information could notably influence the clustering process, leading us to remove all numeric data from the response records. Subsequently, we computed the text similarity among the remaining records based on cosine distance and categorized those with a similarity score above 0.9 within the same service as responses originating from the same reason. In the end, we manually reviewed each category to identify potential new \ipmirror types. Leveraging domain knowledge, these are subsequently transformed into query statements compatible with various \engines. Additionally, regular expressions to match \scanips were formulated and integrated into our pattern database.

\subsection{Behavior monitoring}
Using Flask~\cite{Flask}, we developed an HTTP service in our web honeypot and integrated incorporated fingerprints of router device configuration pages, creating a low-interaction IoT honeypot. We extracted 443 fingerprints from 
devices identified by \engines, and encompassed major router manufacturers such as TPLINK, DLINK, and Tenda. 
% Additionally, we inserted the three forms of access IP into the response header fields and body, as we learned in Section~\ref{sec:preliminary}, to serve as an \ipmirror. \wmy{not sure where to place it, mirror collection or here?@hg}
We also embedded our dynamic links in the default page, which is encoded by a unique visit record as ID, including visiting IP, port, honeypot IP, and timestamp.

In terms of paths, we set common web files including robots.txt, sitemap.xml, and security.txt. Within robots.txt, we defined paths of varying depths that are allowed for crawler access, deliberately including some forbidden paths related to known web vulnerabilities to observe whether \engines would intentionally conduct scans. sitemap.xml provides directory paths of different depths to test the scanning depth and breadth of crawlers. In security.txt, we deliberately included a path in the contact section to observe whether crawlers would read this information. We also embed the unique IDs in the dir of the paths.

Our full-port closed honeypot and popular-port open honeypot are both sniffer honeypots, with the only difference being whether ports are open or not. It utilizes the tcpdump~\cite{tcpdump} to monitor traffic across all ports and combines Berkeley Packet Filter (BPF) syntax for precise packet filtering, enabling effective capture of specific types of traffic.
To eliminate potential interference, we migrated the SSH port to a less commonly used service port and blocked traffic on it. This prevented unnecessary data interactions that could skew results. 
To passively acknowledge TCP packets without actively responding at the application layer, we use \texttt{nc -lk} and \texttt{nc -luk} to listen on TCP and UDP ports, respectively.

\begin{table}[t]
    \centering
    \caption{Default scanning path of \engines in web service scanning.}
    \label{tab:web_path}
    \resizebox{\linewidth}{!}{
    \begin{tabular}{c c c c c}
        \toprule
        \textbf{Path}   & \textbf{Censys}   & \textbf{Shodan}   & \textbf{FOFA} & \textbf{ZoomEye} \\
        \midrule
        /                           & \ding{51} & \ding{51} & \ding{51} & \ding{51} \\
        /robots.txt                 & \ding{55} & \ding{51} & \ding{55} & \ding{51} \\
        /favicon.ico                & \ding{51} & \ding{51} & \ding{51} & \ding{51} \\
        /.well-known/security.txt   & \ding{55} & \ding{51} & \ding{55} & \ding{51} \\
        /sitemap.xml                & \ding{55} & \ding{51} & \ding{55} & \ding{55}\\
        /nice ports,/Trinity.txt.bak & \ding{55} & \ding{55} & \ding{51} & \ding{55}\\
        \bottomrule
    \end{tabular}
    }
\end{table}

\begin{table}[t]
    \centering
    \caption{Sensitive paths scanned by each \engine. ``Ratio'' refers to the percentage of total scanning traffic conducted by the engine that targets the path.}
    \label{tab:sensitive_path}
    \resizebox{\linewidth}{!}{
    \begin{tabular}{c c c c c}
        \toprule
        \textbf{Engine} & \textbf{Type} & \textbf{Path} & \textbf{Request Times} & \textbf{Ratio} \\ 
        \midrule
        \multirow{9}{*}{Censys} & \multirow{9}{*}{\centering Web(Prometheus)} & /api/v1/label/goversion/values & 26,242 & 1.45\% \\

         &  & /api/v1/label/goversion/values & 26,242 & 1.45\% \\ 
         & & /api/v1/query & 26,195 & 1.45\% \\ 
         & & /api/v1/labels & 26,141 & 1.44\% \\ 
         & & /api/v1/label/\_\_name\_\_/values & 26,118 & 1.44\% \\ 
         & & /api/v1/targets & 25,648 & 1.42\% \\ 
         & & /api/v1/label/version/values & 25,619 & 1.42\% \\ 
         & & /api/v1/status/config & 13,015 & 0.72\% \\ 
         & & /tr064dev.xml & 4,927 & 0.27\% \\ 
         & & /api/json & 287 & 0.02\% \\ 
        \midrule
        \multirow{25}{*}{Shodan} & \multirow{25}{*}{\centering IoT(IP Camera)} & /cgi-bin/authLogin.cgi & 5,459 & 1.31\% \\
     & & /filestation/wfm2Login.cgi & 5,099 & 1.22\% \\
     & & /photo & 4,933 & 1.18\% \\
     & & /video & 4,878&1.17\% \\
     & & /snapshot.cgi&750&0.18\% \\
     & & /cgi-bin/viewer/video.jpg&528&0.13\% \\
     & & /cgi-bin/snapshot.cgi&519&0.12\% \\
     & & /snapshot.jpg&485&0.12\% \\
     & & /tmpfs/auto.jpg&465&0.11\% \\
     & & /cgi-bin/view/image&276&0.07\% \\
     & & /axis-cgi/jpg/image.cgi&273&0.07\% \\
     & & /ipcam/jpeg.cgi&272&0.07\% \\
     & & /ISAPI/Streaming/channels/101/picture&268&0.06\% \\
     & & /jpg/image.jpg&266&0.06\% \\
     & & /Streaming/channels/1/picture&265&0.06\% \\
     & & /Streaming/channels/101&261&0.06\% \\
     & & /image/jpeg.cgi & 258 & 0.06\% \\
     & & /img/snapshot.cgi & 253 & 0.06\% \\
     & & /-wvhttp-01-/GetLiveImage & 251 & 0.06\% \\
     & & /-wvhttp-01-/GetOneShot & 250 & 0.06\% \\
     & & /videostream.cgi & 223 & 0.05\% \\
     & & /get\_status.cgi & 219 & 0.05\% \\
     & & /videostream.asf & 218 & 0.05\% \\
     & & /cgi-bin/video\_snapshot.cgi & 217 & 0.05\% \\
     & & /snap.jpg & 212 & 0.05\% \\
        \midrule
        FOFA & Web(Elasticsearch) & /\_cat/indices & 	199 &	0.23\% \\
        \midrule
        \multirow{2}{*}{ZoomEye} & \multirow{2}{*}{IoT(OpenWrt Router)} & /cgi-bin/luci/ &	3,059 &	4.89\% \\
                & & /studio/index.html &	895	& 1.43\% \\
        \bottomrule
    \end{tabular}
    }
\end{table}

\begin{table*}[ht]
    \centering
    \caption{The main User-Agents used by \engines.}
    \label{tab:useragent}
    \resizebox{\linewidth}{!}{
    \begin{tabular}{c c c c}
    \toprule
        \textbf{Engine} & \textbf{Path} & \textbf{User-Agent} & \textbf{Ratio}\\
        \midrule
        \multirow{2}{*}{Censys} & \multirow{2}{*}{\centering all} & Mozilla/5.0 (compatible; CensysInspect/1.1; +https://about.Censys.io/) &69\%  \\
         &  & - &31\%\\
        \midrule
        \multirow{6}{*}{\centering Shodan} & \multirow{1}{*}{/} & Mozilla/5.0 (Windows NT 6.1) AppleWebKit/537.36 (KHTML, like Gecko) Chrome/41.0.2228.0 Safari/537.36 & 31\% \\
        \cmidrule{2-4}
            & \multirow{3}{*}{/favicon.ico} & Mozilla/5.0 (Windows NT 10.0; Win64; x64) AppleWebKit/537.36 (KHTML, like Gecko) Chrome/102.0.5005.63 Safari/537.36 & \multirow{3}{*}{31\%}  \\
            & & Mozilla/5.0 (X11; Linux x86\_64) AppleWebKit/537.36 (KHTML, like Gecko) Chrome/98.0.4758.102 Safari/537.36 &  \\
            & & Mozilla/5.0 (Macintosh; Intel Mac OS X 10.15; rv:80.0) Gecko/20100101 Firefox/80.0 & \\
            \cmidrule{2-4}
            & camera & python-requests &15\% \\
            & other & - &36\%\\
        \midrule
        \multirow{2}{*}{FOFA} & /favicon.ico & Mozilla/5.0 (Windows NT 6.1) AppleWebKit/537.36 (KHTML, like Gecko) Chrome/49.0.2623.112 Safari/537.36 &17\%  \\
        & other & - &81\% \\
        \midrule
        \multirow{2}{*}{ZoomEye} & \multirow{2}{*}{\centering all} & Mozilla/5.0 (X11; Linux x86\_64) AppleWebKit/537.36 (KHTML, like Gecko) Chrome/86.0.4 240.111 Safari/537.36 &29\%  \\
         &  & - &55\% \\
        \bottomrule
    \end{tabular}
    }
\end{table*}

% \section{Sensitive Path}

% \section{Service Identification}
\begin{table*}[h]
    \centering
    \caption{The Multi-port identified protocols and their corresponding number of probe types and the list of ports. Since the number of ports corresponding to fallback probes is vast, their results are not displayed here.}
    \label{tab:service_multiport}
    
    \begin{tabular}{l l l}
        \toprule
        \textbf{Service} & \textbf{\# of Types} & \textbf{Target Ports} \\
        \midrule
            \rowcolor{gray!20}
            Secure Shell & 2 & [22, 2222] \\ 
            Network Basic Input/Output System & 7 & [25, 137, 139, 7587, 11382, 23915, 29844, 31530, 34125, 34303, 40013, 44893, 47415] \\ 
            \rowcolor{gray!20}
            OpenVPN & 2 & [443, 500, 1194] \\ 
            
            Socket Secure & 4 & [1080, 5555, 5678, 7777, 7788, 7890, 8888] \\ 
            \rowcolor{gray!20}
            Microsoft SQL Server &5&\makecell[l]{[427,1433,1434,7025,10001,16592,20748,21429,22637,28864,31980,41372,51668,55010,\\61870]}\\
            
            Mikrotik Router &7&\makecell[l]{[111,2000,4478,7215,8728,10151,23810,24285,27527,32400,38676,40454,41787,49122]}\\
            \rowcolor{gray!20}
            Session Initiation Protocol &6&\makecell[l]{[4871,5060,5061,6060,6788,8320,10325,10326,14396,16319,19867,25841,27650,31492,\\34182,35042,37997,39510,39849,46321,46837,49699,50929,54023,58038]}\\
            
            NAT Port Mapping Protocol &6&\makecell[l]{[69,80,520,1812,1877,2869,3389,3600,3786,5351,5432,6340,6604,6969,\\7001,7320,7398,8000,8290,8835,8945,9999,10690,11211,12205,16397,17180,23205,\\23209,23627,24046,24588,24921,28348,30718,32626,34425,34664,35494,37834,40257,40891,\\41145,41216,41407,45127,45567,46062,47868,51168,51261,53413,53878,54232,57385,58682,\\59478,64738,64940,65501]}\\
            \rowcolor{gray!20}
            X Window System & 4 & [6000, 6001, 6002] \\ 
            
            Redis & 5 & [6379, 6666, 7000] \\ 
            \rowcolor{gray!20}
            Ubiquiti Discovery Protocol &4&\makecell[l]{[19,382,3745,4095,5094,9185,10001,11977,18798,19132,20004,22153,22834,24669,\\27464,32157,32521,32889,34344,36712,38130,39396,39509,42481,44045,47395,51887]}\\
            
            Domain Name System &7&\makecell[l]{[53,69,174,1967,2967,5353,9646,10001,20104,21301,28159,29997,30855,32276,\\37165,47268,48409]}\\
            \rowcolor{gray!20}
            Network Time Protocol &8&\makecell[l]{[123,1632,2112,9577,14983,23708,33270,36503,42507,51759,52315,53075,61172,65037]}\\
            
            X Display Manager Control Protocol &2&\makecell[l]{[69,177,1910,12816,13495,13636,14694,15330,15742,17790,25622,30397,32888,36997,\\38792,40538,45197,47122,50647,59675]}\\
            \rowcolor{gray!20}
            Negotiation of NAT-Traversal in the IKE &1&\makecell[l]{[500,1194,1891,3997,4304,4500,6154,7928,8209,12390,12429,14973,16160,20969,\\22993,24512,25270,26680,28200,31788,33172,34949,34956,38381,38538,40126,40224,40727,\\42850,42910,44568,44806,45708,46061,49109,49147,51822,54015,59491,63038,63284,64367]}\\
            
            Routing Information Protocol & 6 & [520, 2222, 4301, 17948, 23103, 27305, 35315, 35405, 36333, 38527, 64648] \\ 
            \rowcolor{gray!20}
            Universal Plug and Play & 3 & [1474, 1900, 16435, 21721, 24695, 32410, 32414, 37215, 38412, 38599, 45913, 56721] \\ 
            
            Citrix MetaFrame application & 2 & [1604, 23168, 23261, 33352, 38205, 38890, 41912, 46508, 58206, 58344, 58686] \\ 
            \rowcolor{gray!20}
            RADIUS & 2 & [1645, 1812, 6574, 16531, 20899, 26701, 29322, 48794, 52452, 54347] \\ 
            
            Simple Object Access Protocol & 3 & [370, 2191, 3702, 8446, 21229, 35830, 56006] \\ 
            \rowcolor{gray!20}
            Apple Remote Desktop & 4 & [3283, 9334, 13853, 14434, 17847, 43041, 47851, 52327, 55123, 56498, 62279, 63176] \\ 
            
            A2S Query protocol &3&\makecell[l]{[4131,8626,12893,18745,21025,22767,24018,27015,27016,27105,28015,32165,41700,57896]}\\
            \rowcolor{gray!20}
            VxWorks Wind DeBug agents &3&\makecell[l]{[4210,12819,14567,14771,17185,18265,20379,26764,28940,31339,48717,49530,49661,51202,\\57125,57175,57381,57609,62151,63735]}\\
            
            Datagram Transport Layer Security &2&\makecell[l]{[5061,5257,5684,5738,6625,7243,11920,19604,20374,20720,21406,28845,31436,31966,\\33703,38765,39434,39783,50338,52540,52668,52685,53405,59168,63340]}\\
            \rowcolor{gray!20}
            DNS-Based Service Discovery & 2 & [5353, 18235, 18529, 24173, 24626, 25301, 26081, 29939, 45293, 62663, 65337] \\ 
            
            Building Automation and Control Networks & 2 & [5407, 6833, 7642, 9140, 18427, 25337, 31513, 33728, 42168, 47808] \\ 
            \rowcolor{gray!20}
            PC Anywhere & 4 & [5001, 5632, 10522, 31348, 39939, 41650, 42730, 50388, 57664] \\ 
            
            Distributed hash table & 3 & [6881, 13001, 24530, 29579, 29899, 44629, 44633, 47199, 48097, 48688] \\ 
            \rowcolor{gray!20}
            Simple Mail Transfer Protocol & 2 & [25, 587] \\ 
            
            GPRS Tunneling Protocol & 2 & [2123, 2152, 3386] \\ 
            \rowcolor{gray!20}
            Session Traversal Utilities for NAT & 3 & [3478, 8088, 37833] \\ 
            
            Constrained Application Protocol & 2 & [5673, 5683] \\ 
            \rowcolor{gray!20}
            Android Debug Bridge & 2 & [5555, 9001] \\ 
            
            Java Remote Method Invocation & 1 & [6000, 10001] \\ 
            \rowcolor{gray!20}
            Java Debug Wire Protocol & 1 & [8000, 9000] \\ 
            
            All-Seeing Eye & 1 & [8000, 11211] \\ 
        \bottomrule
    \end{tabular}
\end{table*}

\subsection{Web Scanning Strategy}
From 2,503,761 requests our web honeypot captured, we observed that on average, Shodan sent 17.69 requests, Censys sent 60.25 requests, Fofa made 5.87 requests, and ZoomEye sent 3.07 requests per day to each honeypot, revealing distinct patterns. Their path access strategies also differed significantly.
% 
% \noindent \textbf{Handling of URL formats.}
% We noticed that \scanips of ZoomEye sometimes appended an extra slash ('/') to the same path during exploration, for instance, both '/favicon.ico' and '/favicon.ico/' were scanned. This phenomenon suggests that ZoomEye might employ URL normalization strategies during scans. This normalization ensures consistency and accuracy in scan requests, as different web servers might handle paths with and without trailing slashes differently.
ZoomEye's \scanips occasionally appended an extra slash ('/') to paths, such as scanning both '/favicon.ico' and '/favicon.ico/', suggesting a URL normalization strategy to handle trailing slashes across different web servers.

% \noindent \textbf{Distinct scanning phases.} 
We also observed that \engines perform scanning in multiple phases.
Shodan, for example, begins with a broad scan using the User-Agent ``Mozilla/5.0 ... Safari/537.36'' for default paths. Once it discovers information of interest, it launches a focused scan, switches to Python-based User-Agents (such as \texttt{``python-requests/2.27.1''} and \texttt{``python-requests/2.23.0''}) and scans multiple specific paths. This change in User-Agent suggests that Shodan potentially employs diverse strategies and techniques during the scanning process.

\end{document}